\documentclass[sigconf, nonacm, 10pt]{acmart}

\AtBeginDocument{%
  }

\usepackage{graphicx}
\usepackage{subcaption}
\usepackage{booktabs}
\usepackage{array}
\usepackage{enumitem}
\usepackage{bm}

\begin{document}

\title{Beyond the Bulk-Load Honeymoon: Longitudinal Benchmarking of Learned Indexes Under
Sustained Storage Workloads}

\author{Emmilly Immaculate Namuganga}
\affiliation{%
  \institution{MPI-SWS}
  \city{Saarbr\"ucken}
  \country{Germany}}
\email{enamugan@mpi-sws.org}
\author{Laurent Bindschaedler}
\affiliation{%
  \institution{MPI-SWS}
  \city{Saarbr\"ucken}
  \country{Germany}}
\email{bindsch@mpi-sws.org}
\renewcommand{\shortauthors}{Namuganga and Bindschaedler}



\begin{abstract}
Learned indexes are often reported to be much faster than B-trees, but the benchmarks
behind those numbers collapse a whole workload into one average throughput value.
That average mixes the fast early operations, run just after the model was fitted, with
the slower ones that come later.
We call the fast early period the \emph{bulk-load honeymoon}, and a single-shot benchmark
cannot separate it from the throughput an index keeps up over time.
We propose a \emph{longitudinal benchmarking} protocol. It runs an index across many
batches without ever resetting it, and records throughput, tail latency and memory for
every batch.
We run nine indexes on four real-world datasets and one synthetic dataset, with three
seeds per configuration.
On the four real-world datasets LIPP and DILI lose 36 to 45\% of their first-batch
throughput by the last batch, and ALEX loses 25 to 32\%.
Sampled P99 read latency grows by up to 2.2$\times$, and memory per stored key grows
by up to 78\%.
This changes which index looks best. On libio the standard GRE benchmark ranks LIPP
13\% ahead of ARTunsync, but by the end of our run LIPP is 11\% behind it.
Of the nine cases where a learned index starts ahead of ARTunsync, five end behind it.
The decline continues out to 800M keys and 24 batches, and it also appears with 8 to 32
threads, with updates and deletes, at read-heavy and write-heavy mixes, and at three
batch sizes.
Comparing runs that end with the same table but bulk-load different amounts shows that
for LIPP and DILI the loss comes from inserting keys, not from holding more of them.
ALEX barely notices how the keys arrived.
Rebuilding every two batches makes LIPP's last batch on libio 66\% faster, at a
cost of 25 to 106 seconds per rebuild at 200M keys.
\end{abstract}

\maketitle

\section{Introduction}
\label{sec:intro}

Learned indexes replace B-tree internals with machine learning models that approximate the
cumulative distribution function (CDF) of the stored keys~\cite{kraska2018case}.
Many updatable designs followed.
ALEX~\cite{ding2020alex}, LIPP~\cite{wu2021lipp}, and DILI~\cite{li2023dili} show
order-of-magnitude throughput improvements over B-trees on in-memory workloads.
PGM-Index~\cite{ferragina2020pgm}, SALI~\cite{ge2023sali}, and
Hyper~\cite{zhang2024hyper} follow with different design trade-offs.
These designs are now proposed as drop-in replacements for memtable layers, block indexes,
and file-system metadata structures in production key-value stores~\cite{dong2021rocksdb},
with learned indexes already built into LSM-trees~\cite{dai2020bourbon} and a Google-scale
disk-based database~\cite{abulibdeh2020learned}.

This raises a practical question that existing benchmarks do not answer. How does a
learned index perform after a long period of continuous inserts?
A learned index fits its models to the keys it holds when it is built.
As inserts arrive, especially into regions the original models never saw, the index has
to change its structure to make room, and the models it starts with stop describing the
keys well.
Structures such as ART~\cite{leis2013art} and the B-tree~\cite{graefe2011modern} also
change as keys arrive, but they carry no fitted state that can go out of date.
No current benchmark tracks how the cost of this unfolds over time.

Most benchmarks used to justify performance gains report one throughput value per run.
SOSD~\cite{kipf2019sosd} set the protocol for read-only evaluation.
GRE~\cite{wongkham2022updatable} extended it to mixed read and write workloads, and
RoBin~\cite{luo2025understanding} added stress scenarios that catalog failure modes.
Each of them reports numbers that cover a whole run, so fast early operations and slower
later ones are averaged together.
We call the fast early period the \emph{bulk-load honeymoon}, the peak just after the
index is built, which a single number per run cannot separate from sustained throughput.

We propose a \emph{longitudinal benchmarking} protocol that splits execution into
successive fixed-size batches, never resets the index between batches, and measures
throughput, tail latency and memory for every batch over hundreds of millions of
operations. We run nine indexes on four real-world datasets and one synthetic dataset.
Our measurements show three things.

(1)~An index that leads early often does not lead at the end. We compare each learned
index against ARTunsync, the strongest traditional baseline on throughput in our study. Across our
five datasets there are nine cases where a learned index is ahead at the first batch,
seven of them on the four real-world datasets. Five of the nine are behind by the last
batch, one finishes level, and three keep their lead
(Table~\ref{tab:reversal}). On libio all three learned indexes start ahead and all
three end behind.

LIPP is the clearest case. Its throughput falls 44\% over the run, from 4.77 to
2.68\,Mops/s. The effect on index choice is easiest to see against the standard
benchmark. GRE runs this same workload and reports one throughput average, which places
LIPP 13\% ahead of ARTunsync at 3.18 against 2.82\,Mops/s. That average is what a
practitioner would use to choose. Measured per batch, the same workload ends with LIPP
at 2.68 and ARTunsync at 3.00, so LIPP finishes 11\% behind. The average hides the
crossing.

(2)~Where a learned index keeps its lead, the source we read has a matching pattern.
ALEX keeps its lead on COVID, and it is the only one of ALEX, LIPP and DILI whose
self-repair can reach every part of the tree and is never switched off. On the
synthetic uniform dataset both learned indexes keep their lead because
ARTunsync degrades there too.

(3)~Degradation shows up in more than one metric at once. Throughput falls, tail
latency drifts, and memory per key grows. A single number per run shows none of these
trends.

The rest of this paper tests whether that pattern holds under conditions a practitioner
would actually meet. We vary scale up to 800M keys, thread count from 8 to 32, the read
and write mix from read-heavy to write-heavy, the number of operations per batch from 25M
to 100M, the bulk-load fraction, and the operation mix to include updates and deletes.

\noindent\textbf{Contributions.}
\begin{enumerate}[leftmargin=*, topsep=2pt, itemsep=1pt]
  \item A \textbf{longitudinal benchmarking protocol} that runs an index across
        successive fixed-size batches without resetting its state, and reports
        throughput, tail latency, and memory footprint for every batch
        (\S\ref{sec:method}).
  \item An \textbf{empirical study} of nine indexes on four real-world datasets and one
        synthetic dataset (\S\ref{sec:eval}). We vary the read and write mix, the
        number of operations per batch, the thread count, and the operation mix. LIPP and
        DILI lose more throughput than the traditional baselines in every variation we
        ran, and so does SALI against ARTolc at every thread count. Ranking reversals
        appear under all of these variations except the delete mix.
  \item A \textbf{scale and duration extension} to 800M keys and 24 batches, run both
        single-threaded and at 8 threads (\S\ref{sec:scale-800m},
        \S\ref{sec:concurrent-scale}). The decline does not level off within the run,
        and added threads do not offset it.
  \item A \textbf{rebuild methodology} that prices periodic re-indexing rather than
        asking whether it helps (\S\ref{sec:reindex}). Rebuilding recovers most of the
        lost throughput for the worst affected indexes at a measured wall-clock cost,
        and the cost splits into the part any structure pays and the part only fitted
        models pay.
  \item The \textbf{repair rules} of ALEX, LIPP and DILI, copied from their source
        and set next to the measured shapes of decline (\S\ref{sec:mechanisms}).
\end{enumerate}

\section{Motivation}
\label{sec:background}

\subsection{Updatable Learned Indexes and the Bulk-Load Peak}

Updatable learned indexes such as ALEX~\cite{ding2020alex}, LIPP~\cite{wu2021lipp} and
DILI~\cite{li2023dili} organise the keys into a tree of nodes, and each node holds a
linear model that maps a key to an approximate position.
At bulk-load time the models are fitted to the whole initial key set, so a lookup can
predict a position directly and needs only a small local correction.

Inserts change this. ALEX stores keys in gapped arrays and, when a node fills up, it
expands the node or splits it and refits the model.
LIPP places each key at the exact slot its model predicts, and when two keys want the
same slot it creates a child node there, so the tree gets deeper where inserts
collide.
DILI also creates child nodes under conflicts and retrains a leaf when it gets too
expensive.
Each index has a rule for when to rebuild part of itself, and \S\ref{sec:mechanisms}
looks at those rules in detail.
Between rebuilds the structure drifts away from the one the bulk load produced, which is
why the model that gives the lowest lookup cost at $t{=}0$ is not the one running at
$t{=}N$.
ART~\cite{leis2013art} and the B-tree~\cite{graefe2011modern} have no models to fit, so
nothing in them goes out of date in this way, although they still get slower as the
table grows.

\subsection{The Single-Shot Measurement Blind Spot}

Table~\ref{tab:protocols} compares the measurement protocols of the five benchmark suites most
relevant to updatable learned indexes.

\begin{table}[t]
\small
\setlength{\tabcolsep}{4pt}
\caption{Benchmark protocol comparison for updatable learned indexes.}
\label{tab:protocols}
\begin{tabular}{@{}llp{1.6cm}p{2.6cm}@{}}
\toprule
\textbf{Protocol} & \textbf{Workload} & \textbf{State} & \textbf{Output} \\
\midrule
SOSD~\cite{kipf2019sosd}          & read-only & fresh/run  & 1 throughput value \\
GRE~\cite{wongkham2022updatable}  & mixed r/w & fresh/run  & 1 tput + latency \\
Sun et al.~\cite{sun2023learned}  & mixed r/w & fresh + accum. & cost growth, no trajectory \\
RoBin~\cite{luo2025understanding} & mixed r/w & fresh/run  & 1 tput + failure modes \\
NeurBench~\cite{zhao2025neurbench}& mixed r/w & fresh/scenario & snapshot per designed drift \\
\textbf{Ours}                     & mixed r/w & \textbf{accumulated} & \textbf{per-batch tput + degradation + periodic rebuild} \\
\bottomrule
\end{tabular}
\vspace{-3mm}
\end{table}

SOSD~\cite{kipf2019sosd} and the study of Marcus et al.~\cite{marcus2020benchmarking}
established the read-only baseline.
GRE~\cite{wongkham2022updatable} extended it to mixed read/write workloads over ten
updatable structures.
Sun et al.~\cite{sun2023learned} extend the study to nine structures in their TLI testbed.
Their ``block-wise loading'' evaluation is the closest prior work to ours, but results
are reported only for ALEX and only as mean latency per block.
Mean latency can hide tail behaviour. In our runs LIPP's read P99.9 grows 5.5$\times$
over a run at 25M operations per batch (\S\ref{sec:repair-cost}).
We report tail latency and memory for every batch and for every index we evaluate.
RoBin~\cite{luo2025understanding} extends GRE with 124 stress scenarios that catalog
failure modes under extreme stress.
We instead measure the gradual degradation of normal operation.
NeurBench~\cite{zhao2025neurbench} adds workload and distribution drift as explicit
benchmark parameters, but each drift step resets the index to a fresh state. It measures
how an index performs \emph{under} a given distribution, not how it degrades as inserts
accumulate.
Apart from Sun et al.'s single block-wise ALEX result, every protocol above resets index
state between runs.
Bindschaedler et al.~\cite{bindschaedler2021towards} argued that a learned component
cannot be judged on a single average score, and asked for benchmarks that report both
descriptive statistics and behaviour over time.
The suites that came after go part of the way.
GRE reports tail latency next to mean throughput, and RoBin catalogues the conditions
under which an index fails, so the descriptive statistics are there.
The behaviour over time is not.
Each of those numbers still covers a whole run, so none of them shows how an index
changes while it runs.

Our protocol measures that change, and it has two parts.

The first is how the benchmark runs.
We split the workload into fixed-size batches, carry index state across them, and draw
inserts from a pool that must be fully consumed (\S\ref{sec:method}), so the result is
one measurement per batch rather than one per run.

The second is three measurements that explain the degradation rather than just record
it, none of which a prior learned-index benchmark reports.
Varying the bulk-load fraction asks how much of the degradation was caused by inserting
rather than by holding more keys.
Several runs are grown to nearly the same final table size from different starting
points, so they differ mainly in how many keys arrived by insert, and the throughput gap
between them cannot be blamed on table size (\S\ref{sec:path-dependence}).
Periodic rebuild asks what removing that accumulated cost is worth.
The harness discards and rebuilds the live index every $k$ batches and records the pause
(\S\ref{sec:reindex}).
The cost breakdown asks how much of that pause is specific to learned indexes.
It splits the pause into the snapshot and sort that any structure pays and the model
fitting that only a learned index pays (\S\ref{sec:reindex}).

\section{Longitudinal Benchmarking Protocol}
\label{sec:method}

The protocol is built on the GRE benchmark code~\cite{wongkham2022updatable} and changes
two things about it.

The first is the execution model.
GRE runs a workload as a single burst and reports one number.
We split the same workload into a sequence of fixed-size batches, report one number per
batch, and let the index keep its state across batch boundaries.

The second is latency accounting.
GRE pools every sampled operation into a single latency distribution.
We record read and write latency separately, because the two move for different reasons
and at different rates.
A pooled distribution would have hidden the clearest mechanism result in this paper,
which is ALEX holding its read latency flat while its write latency spikes by
$33\times$ (\S\ref{sec:repair-cost}).

Timing otherwise follows GRE, using TSC sampling of a configurable fraction of
operations, 1\% by default.
We run at $\rho_\text{init}{=}0.25$ with a 50/50 read and insert mix and shuffled
operation order.
The single-shot GRE baseline we compare against in \S\ref{sec:honeymoon} uses these same
settings and the same seeds, so the two differ only in whether execution is split into
batches.

\noindent\textbf{Setup.}
Each dataset provides $N = 200$\,M keys.
An index bulk-loads $\rho_\text{init} N = 50$\,M of them.
The remaining 150\,M keys form the insert pool, and the batches that follow must consume
all of it.

\noindent\textbf{Batch execution.}
Each batch runs $M = 50$\,M operations.
Reads, and updates or scans where the workload includes them, are sampled uniformly from
the keys already in the index, taken before that batch's own inserts join it.
Inserts are drawn in order from the remaining pool.
The batch is then shuffled, so operation types are interleaved rather than grouped.
The number of batches is
$B = \lfloor (1{-}\rho_\text{init})N / (M \cdot r_\text{insert}) \rfloor$,
which is what makes the pool run out at the last batch, and $B$ grows as the write
fraction shrinks.
Between batches the index is never rebuilt or reset, unless a rebuild experiment says
otherwise (\S\ref{sec:reindex}).
It carries forward everything it has accumulated, including model error, split history,
and memory layout.
All operations for all batches are generated once, before any index runs, from that
run's seed.
Every index then replays that same sequence, so a difference between two indexes is a
difference in behaviour rather than in which keys they happened to see.

\noindent\textbf{Default workload.}
The default run sits at GRE's balanced point, $r_\text{read} = r_\text{insert} = 0.5$.
That gives $B{=}6$ batches, each adding 25\,M keys, so the table grows from 50\,M to
200\,M keys over the run.
This is the configuration behind every result in \S\ref{sec:eval} unless stated
otherwise.

\noindent\textbf{Seeds and repetition.}
Every configuration in this paper is run three times, with seeds 1866, 5 and 72.
We report the mean across the three, and the range across them where a figure shows a
band.
The seed decides which keys are sampled for reads and how the batch is shuffled, so
changing it gives a different operation sequence, and within one seed every index still
replays the same sequence as every other.

\noindent\textbf{Metrics.}
\label{sec:metrics}
We record four things per batch.
Aggregate throughput in Mops/s, sampled read latency at P\textsubscript{50},
P\textsubscript{99} and P\textsubscript{99.9}, sampled write latency at
P\textsubscript{50} and P\textsubscript{99}, and the memory footprint each index reports
for its own nodes and data.
From those series we compute four summary statistics.
\emph{Retention} is the last batch over the first.
\emph{Trend} is the mean of the last two batches over the mean of the first two, which is
less sensitive to one batch swinging.
\emph{Spread} is the residual coefficient of variation around a Theil--Sen trend line,
which is batch-to-batch volatility with the trend removed.
\emph{Latency drift} is P\textsubscript{99} at the last batch over P\textsubscript{99} at
the first.

\section{Evaluation}
\label{sec:eval}

\subsection{Experiment Settings}

\noindent\textbf{Testbed.}
All experiments run on a dual-socket Intel Xeon Gold 6134M server (8 cores per socket,
3.20\,GHz, 25\,MB L3) with 768\,GiB DRAM.
Index state resides entirely in main memory.
Single-threaded experiments are pinned to one core.
Concurrent experiments use 8, 16, 24 or 32 threads on the same machine. The machine has
16 physical cores, so the 24 and 32 thread runs use more threads than physical cores.

\noindent\textbf{Indexes and datasets.}
The single-threaded runs use ALEX, LIPP and DILI (learned), plus ARTunsync and BTree
(traditional). DILI is libio and Planet only, because OSM and the 800M files overflow
its signed 64-bit keys and COVID segfaults in bulk load.
The concurrent runs use AlexOL and SALI~\cite{ge2023sali} (learned), plus
ARTolc and BTrOLC (traditional). The GRE wrapper caps an AlexOL data node at 512\,KB
(32{,}768 entries) but leaves ALEX at its default of 16\,MB, so the two repair in
units that differ by up to 32$\times$. We also ran FINEdex. It is 46 to 70\% slower
than ARTolc at every batch, never leads, and some runs failed the harness success-flag
check, so we leave it out. That check does not read the keys back. The AlexOL numbers
here come from the released GRE binary. Our companion paper reads every key
after a run and finds defects in that binary that lose keys when a root
expansion fires early, and one or two lost keys in some FINEdex
runs~\cite{namuganga2026synchronized}. The companion
paper's own ALEX-OL measurements use a patched binary, so an ALEX-OL number there
is not from the same binary as the AlexOL numbers here.
The real-world datasets are COVID-19 case IDs (easy, left-heavy CDF), libio library release keys
(moderate, near-uniform), OSM node IDs (hard, step-bimodal), and Planet GPS encodings (hard,
extreme right-skew), plus synthetic uniform. The 800M run adds SOSD books.
Figures show libio and Planet, which span the moderate-to-hard range.

\subsection{What Single-Shot Measurement Hides}
\label{sec:honeymoon}

\begin{figure*}[!t]
  \centering
  \includegraphics[width=\textwidth]{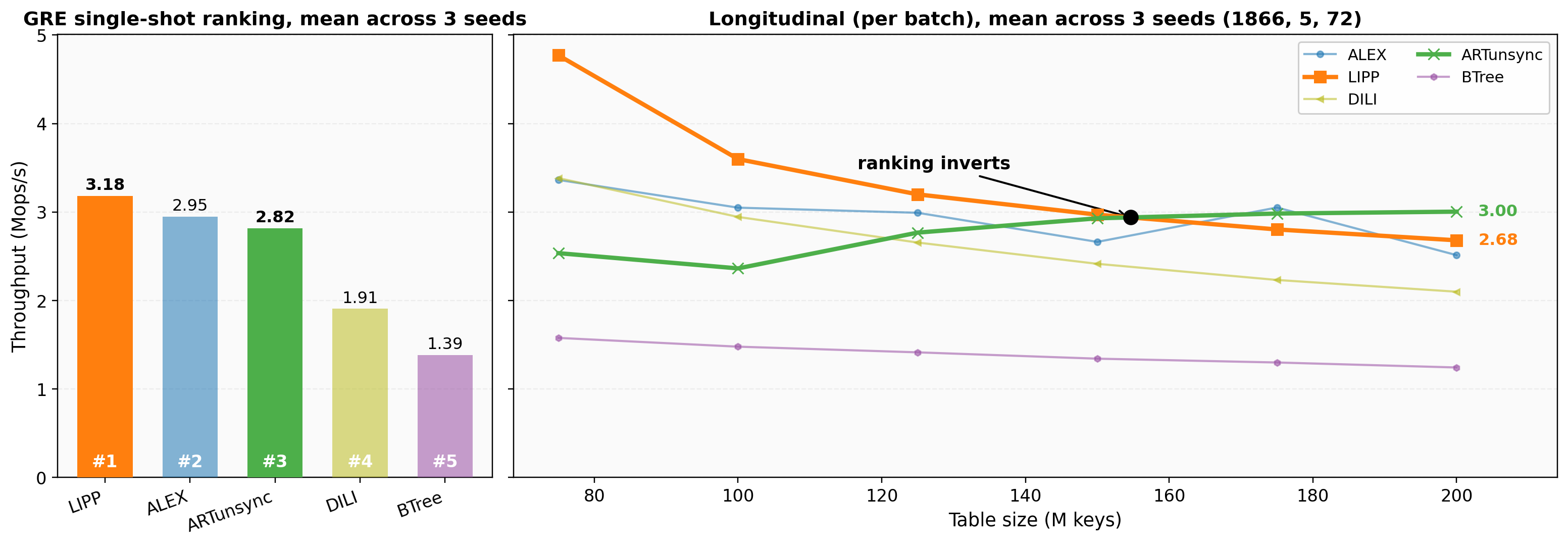}
  \Description{Two panels. Left, a bar chart of GRE single-shot throughput on libio ranking LIPP first and ARTunsync third. Right, longitudinal per-batch curves showing ARTunsync overtaking LIPP near 155M keys and finishing ahead.}
  \vspace{-3mm}
  \caption{The single-shot ranking picks the wrong index on libio, mean across 3 seeds
           (1866, 5, 72). GRE single-shot throughput (left) ranks LIPP first, 13\% ahead
           of ARTunsync in third. Per-batch throughput under sustained load (right)
           inverts that ranking between 150 and 155\,M keys, and LIPP finishes 11\%
           behind (2.68 against 3.00\,Mops/s). On Planet ARTunsync is already ahead of
           LIPP in the GRE single-shot run (2.20 against 2.00\,Mops/s).}
  \label{fig:gre-vs-lon-timeseries}
\end{figure*}

\textbf{Key result:} LIPP's throughput falls 44\% over the run on libio, from 4.77 to
2.68\,Mops/s, closing at 56\% of its first-batch value.
Single-shot measurement misses that decline entirely.
It reports one number, 3.18\,Mops/s, which overstates the throughput LIPP actually
sustains by the end of the run by 19\%.
Figure~\ref{fig:gre-vs-lon-timeseries} shows what that means for an operator choosing an
index. The GRE ranking inverts under sustained load.
The flip is not specific to LIPP. A flip needs both a decline and a lead to lose.
In the GRE ranking DILI has no lead over ARTunsync (1.91 against 2.82\,Mops/s), so even
though it degrades almost as much as LIPP (Table~\ref{tab:metrics}) its GRE ranking
does not flip. SALI shows the same kind of flip against ARTolc at every thread count we
test on libio (\S\ref{sec:concurrent}).

\begin{table*}[t]
\small
\setlength{\tabcolsep}{5pt}
\caption{Whether a learned index that leads at the first batch still leads at the last,
  measured against ARTunsync on all five datasets (mean across seeds 1866, 5 and 72).
  Each pair gives the margin at the first batch and at the last, so $+$88 and $-$11 means
  the index opened 88\% faster than ARTunsync and closed 11\% slower.
  \textbf{Bold} marks a reversal, a lead at the first batch that is gone by the last.
  Nine of the twelve pairs lead at the first batch and five of those reverse. COVID LIPP
  finishes level (its three seeds give $-$2\%, $-$0\% and $+$0\%) and is counted a tie,
  not a reversal. Planet ALEX and both indexes on OSM never lead, because ARTunsync is
  already fastest at the first batch there. DILI was run on libio and Planet only.}
\label{tab:reversal}
\begin{tabular}{@{}l rr rr rr rr rr@{}}
\toprule
 & \multicolumn{2}{c}{\textbf{libio}} & \multicolumn{2}{c}{\textbf{Planet}}
 & \multicolumn{2}{c}{\textbf{COVID}} & \multicolumn{2}{c}{\textbf{OSM}}
 & \multicolumn{2}{c}{\textbf{uniform}} \\
\cmidrule(lr){2-3}\cmidrule(lr){4-5}\cmidrule(lr){6-7}\cmidrule(lr){8-9}\cmidrule(lr){10-11}
\textbf{Index} & First & Last & First & Last & First & Last & First & Last & First & Last \\
\midrule
ALEX & \textbf{$+$33} & \textbf{$-$16} & $-$10 & $-$26 & $+$65 & $+$51 & $-$19 & $-$24 & $+$75 & $+$84 \\
LIPP & \textbf{$+$88} & \textbf{$-$11} & \textbf{$+$12} & \textbf{$-$24} & $+$43 & $-$1 & $-$16 & $-$34 & $+$45 & $+$24 \\
DILI & \textbf{$+$33} & \textbf{$-$30} & \textbf{$+$13} & \textbf{$-$17} & --  & --  & --  & --  & --  & --  \\
\midrule
\textit{outcome} & \multicolumn{2}{c}{\textit{3 reversals}} & \multicolumn{2}{c}{\textit{2 reversals}}
 & \multicolumn{2}{c}{\textit{1 keeps, 1 tie}} & \multicolumn{2}{c}{\textit{no leads}}
 & \multicolumn{2}{c}{\textit{2 keep}} \\
\bottomrule
\end{tabular}
\vspace{-3mm}
\end{table*}

\textbf{How often does the ranking actually invert?}
Table~\ref{tab:reversal} answers that for every pair we ran, against
ARTunsync as the baseline.
Nine of the twelve index and dataset pairs have a lead at the first batch, seven of them
on the four real-world datasets. Five of the nine are behind by the last batch, one
finishes level, and three keep their lead. The remaining three never lead at all.
The margins swing by a lot. LIPP goes from 88\% ahead to 11\% behind on libio, and
DILI from 33\% ahead to 30\% behind.
OSM contributes no cases because ARTunsync is already fastest at the first batch there.

\textbf{A reversal needs two things, and the uniform control supplies only one.}
On uniform both learned indexes lead at the first batch and both still lead at the last,
even though LIPP degrades exactly as much there as on libio (retention 0.56 on both,
Table~\ref{tab:metrics}).
The difference is the baseline. ARTunsync degrades on uniform too, to its worst retention
anywhere (0.65), so it never catches up.
In our data a ranking inverts only when a learned index degrades \emph{and} the
structure it is measured against holds up better, which is why retention alone cannot
predict a reversal and why we report the comparison separately from the decline.
COVID's ALEX is the other index that keeps its lead. It still degrades (retention 0.73),
but its lead at the first batch is large enough to survive.

\label{sec:trajectories}
\textbf{The reversal is one consequence of a broader effect.} Every index with a
fitted model loses throughput across the run, and it loses it in more than one dimension
at once.
Appendix~\ref{app:trajectories} gives the per-batch throughput, read tail latency and
memory trajectories, and Appendix~\ref{app:stats} the four summary statistics for every
index on all five datasets. Three results carry the rest of the paper.

\textbf{BTree sets a growth floor, and it barely moves.} With no model to go stale it
retains 0.75--0.79 and drifts $1.2\times$ at the read tail at the default mix, synthetic
and real alike. That narrow band is what table growth alone costs an index, and it is
the line the others are read against.

\textbf{Every learned index falls below that floor at the default mix.} LIPP retains
0.55--0.61, DILI 0.62--0.64 and ALEX 0.68--0.75. The excess below the floor, not the
decline itself, is the honeymoon cost. LIPP pays it on the synthetic uniform dataset as
much as on a hard real one, so an irregular key distribution is not needed to cause it.

\textbf{Degradation shows up in throughput, latency and memory together.} On libio
LIPP's throughput falls 44\% while its P99 read latency drifts $2.2\times$ and its memory
grows $3.0\times$. DILI's cost per stored key rises 78\%, from 23.9 to 42.5 bytes, while
ARTunsync's \emph{falls} 20\% as its trie shares more prefixes. No single-shot benchmark
reports any of these trends.

\subsection{Finding 1. The degradation belongs to the index, not to one setting}
\label{sec:eval-extension}

The obvious objection to \S\ref{sec:honeymoon} is that we chose the workload, the scale
and the protocol constants, so the reversal may be an artifact of those choices.
We tested seven of them, one at a time, holding everything else at its default and
running three seeds each. Table~\ref{tab:robustness} gives the result.

\textbf{None of the seven makes the effect go away.} In every single-threaded variation
LIPP, and DILI where we ran it, keep less of their first-batch throughput than the
traditional baselines, and under concurrency SALI keeps less than ARTolc at every thread
count on every dataset. Every variation except the delete mix also has cases where a
learned index starts ahead of the traditional baseline and ends behind it. The delete mix
has no such case because its only traditional index, BTree, is the slowest index from
the first batch. Not every lead reverses. Some leads survive, for example on the uniform
dataset and for AlexOL on COVID at 8 and 16 threads, and on the hardest dataset and mix
combinations the learned index never leads at all.
Two of the axes say more than this and are taken up below as findings of their own. The
rest are detailed in the appendix.

\textbf{Concurrency changes when the crossing happens.} On libio the crossing is already
present at 8 threads. On COVID AlexOL keeps its lead at 8 and 16 threads and loses it
only at 24 and 32, because ARTolc gains more from extra threads than AlexOL does
(App.~\ref{app:concurrency}).

\begin{table*}[t]
\small
\setlength{\tabcolsep}{6pt}
\caption{What happens to the degradation and the ranking reversal when one thing is
  varied at a time. Every entry is a mean over seeds 1866, 5 and 72 on the longitudinal
  protocol of \S\ref{sec:method}, with every other parameter held at its default.}
\label{tab:robustness}
\begin{tabular}{@{}p{2.5cm}p{3.6cm}p{10.6cm}@{}}
\toprule
\textbf{Varied} & \textbf{Range} & \textbf{Effect} \\
\midrule
Scale and duration & 200M to 800M keys, 6 to 24 batches &
  No plateau within 24 batches. LIPP's lowest throughput is at the last batch on both
  datasets, and on books it starts ahead of ARTunsync and ends behind it
  (\S\ref{sec:scale-800m}). \\
\addlinespace
Concurrency & 8 to 32 threads, 4 datasets &
  Nine clear leads at the first batch, seven of them reversed, plus one case that opens
  level and ends 18\% behind. Added threads move the crossing earlier, because ARTolc
  gains more from them than the learned indexes do (App.~\ref{app:concurrency}). \\
\addlinespace
Read and write mix & 0.8/0.2 to 0.2/0.8 &
  Reversal at both extremes. On libio a 90\% lead becomes a 9\% deficit read-heavy, and a
  68\% lead becomes an 18\% deficit in only three batches write-heavy
  (App.~\ref{app:spectrum}). \\
\addlinespace
Dataset & 5, including a synthetic uniform dataset &
  LIPP and ALEX are below the floor on all five, DILI on both we ran. On uniform both
  learned indexes keep their lead because ARTunsync degrades too
  (App.~\ref{app:stats}). \\
\addlinespace
Batch granularity & 25M, 50M, 100M operations per batch &
  The same crossing at every size. LIPP enters first and never finishes first,
  ARTunsync enters fourth and finishes first or second (App.~\ref{app:batchsize}). \\
\addlinespace
Bulk-load fraction & $\rho_\text{init}$ of 0.10, 0.25, 0.50 &
  At the final table size LIPP runs 82\% and DILI 92\% faster when half the keys were
  bulk-loaded rather than a tenth. ARTunsync and BTree vary by 1 to 3\%
  (App.~\ref{app:rho}). \\
\addlinespace
Operation mix & read/insert/update and read/insert/delete, 40/40/20 &
  With updates, LIPP starts ahead of ARTunsync and ends behind it on both datasets. With
  deletes, LIPP and DILI still lose more than BTree, and LIPP is overtaken by ALEX
  (App.~\ref{app:mixedops}). \\
\bottomrule
\end{tabular}
\vspace{-3mm}
\end{table*}

\subsection{Finding 2. The cost is paid mainly for inserting, not for holding more keys}
\label{sec:path-dependence}

Every experiment so far compares an index against itself earlier in a run, or against a
different index. In both cases elapsed inserts and table size move together, so neither
separates the two. One experiment does separate them.

We keep the dataset at 200\,M keys and vary only the bulk-load fraction, so
$\rho_\text{init} \in \{0.10, 0.25, 0.50\}$ bulk-loads 20\,M, 50\,M or 100\,M keys and
the rest arrive as inserts. The 0.25 and 0.50 runs end holding all 200\,M keys. The 0.10
run ends at 195\,M, because its 180\,M insert pool does not divide evenly into whole
batches. A smaller table should be faster, so this small difference, if anything,
favours the 0.10 run.
Retention cannot be compared across the runs, because each run covers a different size
range, so we compare throughput at the end of each run instead.

\textbf{Nearly the same table, built two ways, differs by nearly a factor of two.}
At the end of the run on libio, LIPP runs at 1.87\,Mops/s when 90\% of the keys arrived
by insert and 3.40 when only 50\% did, a spread of 82\%. DILI spreads 92\%.
The structures with no fitted model barely notice. ARTunsync spreads 3\% and BTree 1\%,
and ALEX spreads 8\%. Planet gives the same ordering
(App.~\ref{app:rho}).

The final tables are nearly the same size and hold almost the same keys. Mainly the
history differs, and LIPP and DILI end up running roughly half as fast. Periodic
rebuilding points the same way (\S\ref{sec:reindex}). Throwing the history away
recovers most of the lost throughput, and the two indexes it helps most are the same
two.
It also limits the honeymoon framing. For LIPP and DILI a larger bulk load does not just
postpone the decline. The run that bulk-loaded more is faster at every table size where
the runs overlap, so loading more data up front is a lasting benefit on the measured
curve. ALEX is the other case. Its 8\% spread says the final rate barely changes with
how the keys arrived. The expansion schedule puts the first burst at $N_0 \cdot 8/7$,
so loading 100M keys instead of 20M moves that point from 22.9M keys to
114.3M keys~\cite{namuganga2026synchronized}.

\subsection{Finding 3. No plateau within 24 batches}
\label{sec:scale-800m}

\begin{figure*}[!t]
  \centering
  \includegraphics[width=\textwidth]{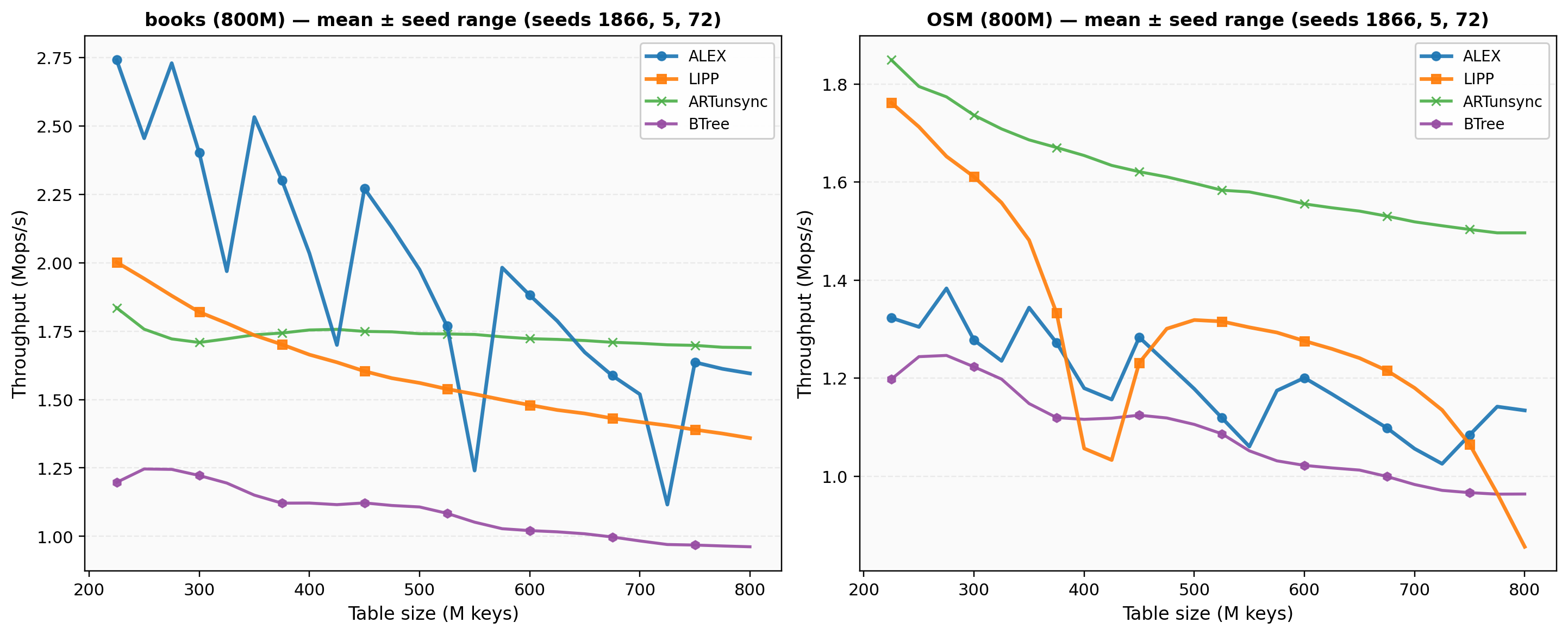}
  \Description{Two line charts showing mean throughput across three seeds (1866, 5, 72)
  vs table size from 200M to 800M keys, for ALEX, LIPP, ARTunsync, and BTree on the
  books and OSM datasets.}
  \vspace{-3mm}
  \caption{Per-batch throughput from 200M to 800M keys (24 batches, single-threaded), mean
           across three seeds (1866, 5, 72), on \textbf{books} (left)
           and \textbf{OSM} (right). The spread across seeds stays small for every index
           over the whole range (at most 6\% of the mean).
           LIPP falls at every batch on books. On OSM it drops sharply around 400M keys,
           partly recovers, and then falls again to its lowest point at the last batch
           (1.76 to 0.86\,Mops/s, against ARTunsync's 1.85 to 1.50).
           ARTunsync and BTree also decline over the run at this scale, more gently than
           LIPP.
           ALEX shows a repeated drop and recover sawtooth, with troughs in the batches
           the expansion schedule names (\S\ref{sec:repair-loop}).}
  \label{fig:scale-800m}
\end{figure*}

Six batches may simply be too short a window to see an index settle. If the decline is a
transient that flattens once the model has absorbed enough inserts, an operator can wait
it out. We tested that by extending to SOSD's largest standard variant, 800\,M keys,
bulk-loading 200\,M and running 24 batches instead of 6 (Figure~\ref{fig:scale-800m}).

\textbf{We see no settling point within 24 batches.} LIPP's lowest throughput of the
entire run falls at the very last batch on both datasets, after 600\,M keys of growth,
and on OSM its deficit against ARTunsync is largest at the last batch. On books LIPP
starts 9\% ahead of ARTunsync and falls behind it after about six batches, so the
reversal also appears at this scale.

\textbf{The longer window also shows that the traditional baselines decline.}
ARTunsync and BTree lose throughput out to 800\,M too (retention 0.80 to 0.92), more
gently than LIPP. So part of any decline is ordinary growth cost, which is why we read
the learned indexes against the BTree growth floor rather than against zero.

\textbf{And the shapes differ.} ALEX oscillates in a repeated drop and recover sawtooth.
On books its troughs sit in the batches that contain 229, 305, 406, 542 and 722\,M
keys, which is where the bulk-loaded data nodes of a 200\,M-key bulk load are predicted
to expand together (\S\ref{sec:repair-loop}).
LIPP declines smoothly on books. On OSM it has one sharp drop and partial recovery near
400M keys before declining again. ALEX's sawtooth is already visible at 200M keys
(Figure~\ref{fig:sawtooth}). The longer run adds more cycles of it, and it is the first
place LIPP's smooth decline breaks.
\subsection{Finding 4. The loss is recoverable, at a measured price}
\label{sec:reindex}

\begin{figure*}[!t]
  \centering
  \includegraphics[width=\textwidth]{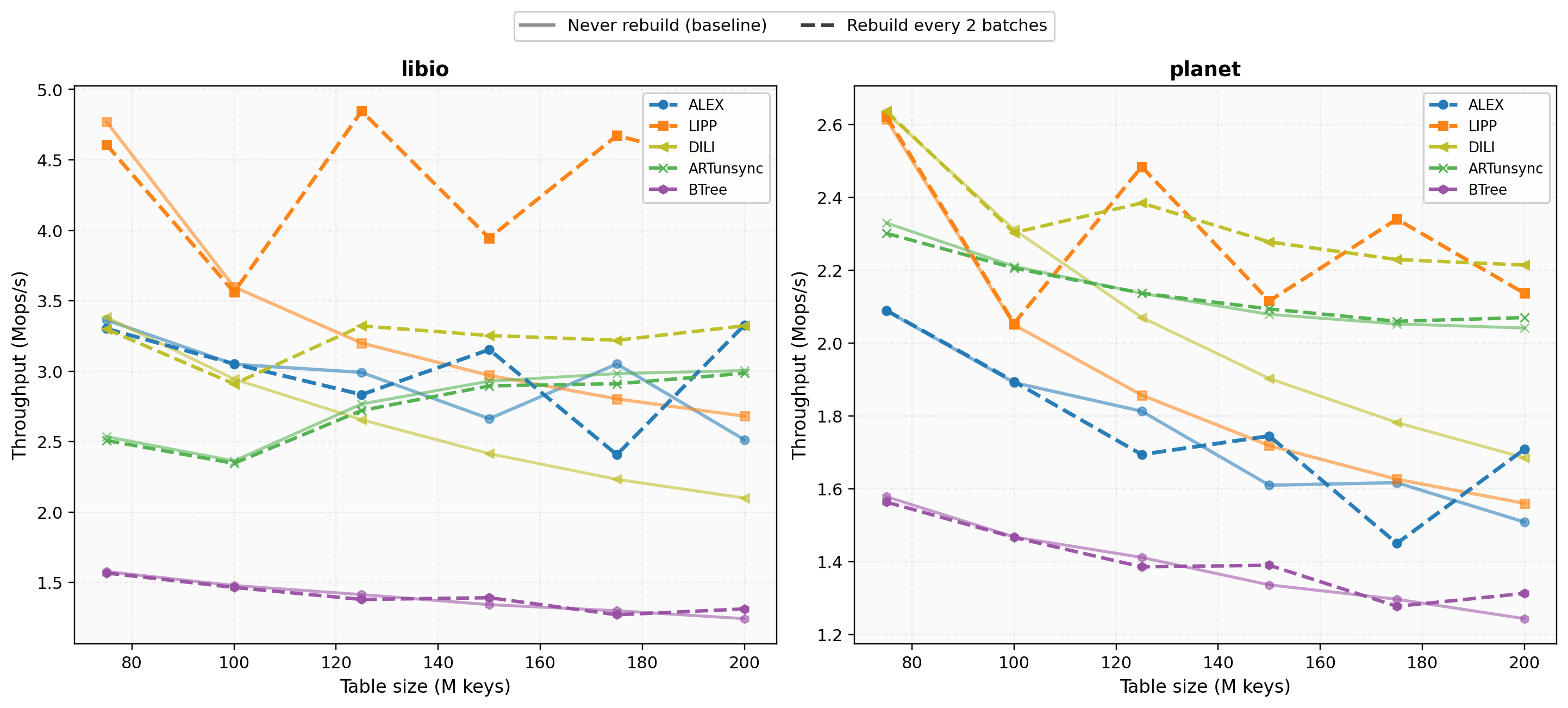}
  \Description{Two line charts comparing per-batch throughput under a never-rebuild
  baseline against a policy that rebuilds every two batches, for ALEX, LIPP, DILI,
  ARTunsync, and BTree on libio and planet.}
  \vspace{-3mm}
  \caption{Per-batch throughput under the never-rebuild baseline (solid) vs.\ a policy
           that discards and freshly bulk-loads the index every two batches (dashed,
           harness flag \texttt{--rebuild\_every=2}), on \textbf{libio} (left) and
           \textbf{planet} (right).
           Both conditions are mean across seeds 1866, 5, and 72.
           Periodic rebuilding recovers most of LIPP's and DILI's lost throughput, with
           final-batch gains of +66\% and +58\% on libio and +37\% and +31\% on planet,
           while ARTunsync and BTree, which have little staleness to recover, gain
           between $-$1\% and +6\% from the same policy.}
  \label{fig:rebuild-policy}
\end{figure*}

\begin{figure}[t]
  \centering
  \includegraphics[width=\linewidth]{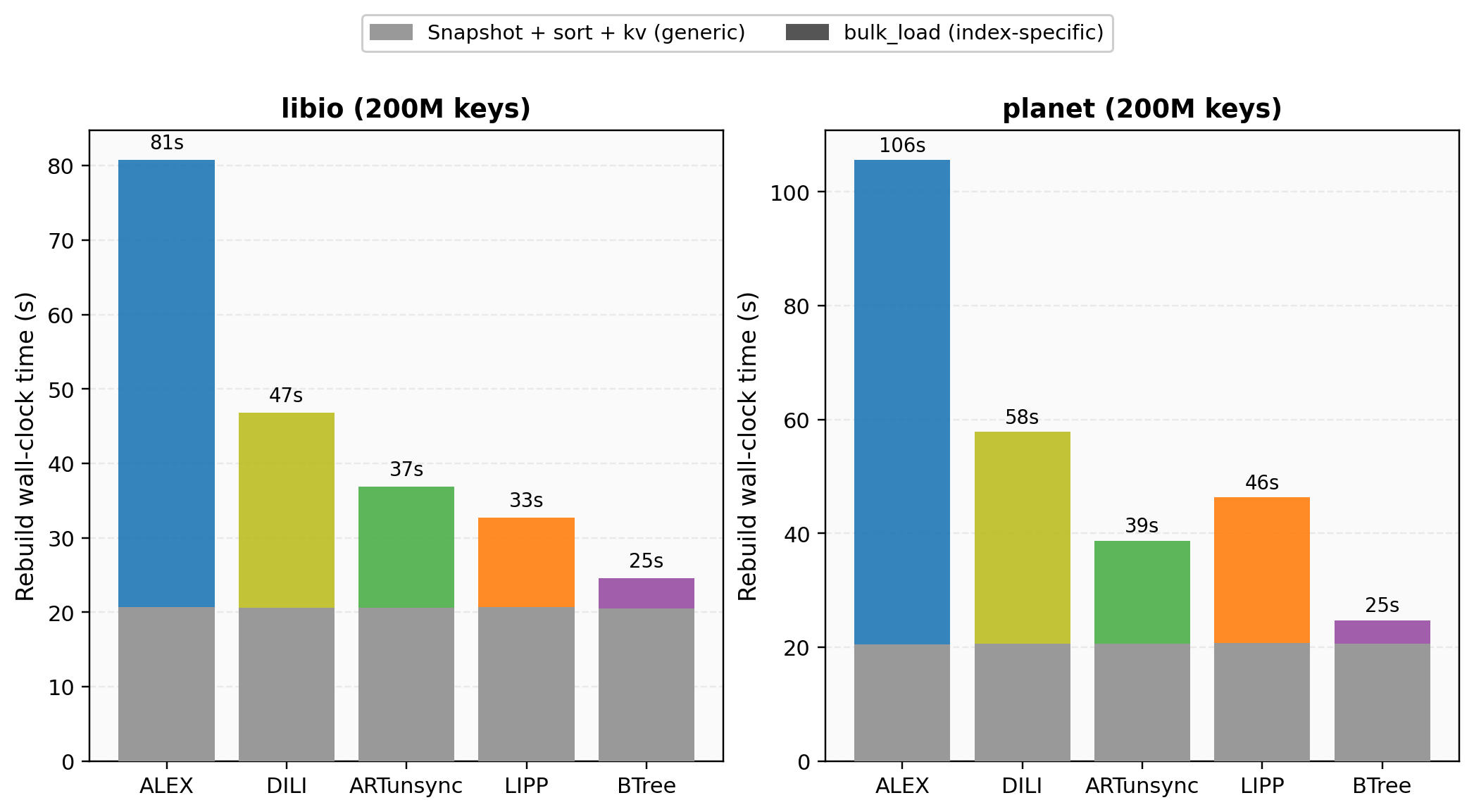}
  \Description{Stacked bar chart showing rebuild wall-clock time at 200M keys for
  ALEX, DILI, ARTunsync, LIPP, and BTree on libio and planet, each bar split into a
  shared generic-overhead segment and an index-specific bulk_load segment.}
  \vspace{-3mm}
  \caption{Full-rebuild wall-clock cost at 200M keys (mean across seeds 1866, 5, and
           72), split into the \textbf{generic} snapshot-extraction, sort, and
           key-value-packing steps every rebuild pays regardless of which index
           consumes them ($\approx$20.6s, identical across indexes and datasets to
           within measurement noise) and the \textbf{index-specific}
           \texttt{bulk\_load} call stacked on top.
           ALEX is priciest overall (81s libio, 106s planet), followed by DILI
           (47s/58s), ARTunsync (37s/39s), and LIPP (33s/46s), and BTree is cheapest
           (25s/25s).
           Because the generic segment is identical for every index, it compresses
           the true end-to-end range to only 3.3$\times$ (libio) / 4.3$\times$
           (planet) between cheapest and priciest, far narrower than the
           14.8$\times$/20.9$\times$ range implied by \texttt{bulk\_load} time alone.
           Rebuild cost remains dataset-hardness-dependent for the globally-fitted
           indexes (ALEX/DILI/LIPP total cost 24--42\% more on planet than libio at
           the same key count) but not for BTree (+0.5\%) or ARTunsync (+5\%).}
  \label{fig:rebuild-cost}
\end{figure}

A natural objection to everything above is that an operator would simply re-index
periodically, so the honeymoon never matters in practice.
It is the right objection. Our answer is that re-indexing works, that it helps most the
indexes Finding~2 points to, and that it costs enough to be a scheduling decision rather
than a free fix.
We answer it with a simulated policy run over a whole workload rather than a single
before and after snapshot, using one addition to the harness that we describe here
because this is the only place it is used.

\noindent\textbf{Periodic re-indexing.}
Finding~2 measures what inserting has cost by the end of a run.
To ask instead whether an operator could rebuild that cost away during normal operation,
we add an optional rebuild every $k$ batches.
After the designated batch the harness snapshots the keys currently in the index,
discards the index, and builds a new one with the competitor's own \texttt{bulk\_load}
on that snapshot.
Later batches continue on the rebuilt index using the same pre-generated operation
sequence as the baseline, so the two runs stay comparable.
The rebuild is timed in four parts (snapshot, sort, packing the key and value pairs,
and \texttt{bulk\_load}) and logged separately, which keeps it out of the per-batch
throughput numbers.
This is separate from the maintenance each index already performs on its own insert
path, such as node expansions and splits (\S\ref{sec:background}), which runs
throughout every batch in both conditions.

\noindent\textbf{Setup.}
We set $k{=}2$ on the balanced growth-only workload, with $N{=}200$\,M keys,
$\rho_\text{init}{=}0.25$ and $M{=}50$\,M operations per batch, so rebuilds fire after
batches 2, 4 and 6 and snapshot 100\,M, 150\,M and 200\,M keys.
We run ALEX, LIPP, DILI, ARTunsync and BTree on libio and planet, single-threaded and
one index per process, over seeds 1866, 5 and 72.
Baseline and rebuild conditions share the same operation traces.
Figures~\ref{fig:rebuild-policy} and~\ref{fig:rebuild-cost} report the resulting
trajectories and the rebuild's own wall-clock cost.

On libio, rebuilding every two batches brings the learned indexes' retention back to
about 1 (0.96 to 1.01). On planet it brings them to 0.81 to 0.84, about the same level
as BTree (0.84). The traditional indexes change little in either case. But the rebuild pauses are
large, from 25\,s for the cheapest index to 106\,s for the most expensive at 200M keys
(Figure~\ref{fig:rebuild-cost}), so re-indexing is an operator trade-off rather than a
free fix.

\textbf{Most of the rebuild cost is not specific to the index.} The snapshot, sort and
packing steps cost the same no matter which index the sorted array goes into. Only
\texttt{bulk\_load} reflects what the index itself does with that data. For BTree these
shared steps are 83 to 84\% of its total rebuild cost (Figure~\ref{fig:rebuild-cost}),
so rebuilding a structure with no model to fit is mostly the cost of re-sorting the
keys. For ALEX the same steps are only 19 to 26\%, because its own \texttt{bulk\_load}
dominates.

\textbf{Rebuilding helps most where inserting hurt most.} LIPP and DILI, the two indexes
that depended most on how the table was built in Finding~2, are also the two that gain
most from periodic rebuilding, by 66\% and 58\% on libio and 37\% and 31\% on planet.
The two experiments are independent. One varies how the keys arrived and the other
throws the history away, and they pick out the same pair. ARTunsync and BTree change by
between $-$1\% and +6\%. ALEX falls in between (+32\% on libio, +13\% on planet).

\section{Repair Rules and the Shape of the Decline}
\label{sec:mechanisms}

The results so far show \emph{how much} each index degrades. This section records the
shape of that decline, and the repair rule each of ALEX, LIPP and DILI uses, taken
from its source.

\subsection{Three Shapes of Degradation}
\label{sec:shapes}

Across every experiment in this paper, the indexes fall into three groups by the
\emph{shape} of their decline, not just its size.

\noindent\textbf{Smooth and steady.} LIPP and DILI lose throughput steadily, batch after
batch. LIPP retains 0.55--0.61 of its first-batch throughput and DILI 0.62--0.64
(\S\ref{sec:stats}), both below the BTree growth floor, LIPP on every dataset and DILI
on the two we ran, and at every batch size we tried. The one exception to the smooth shape is LIPP at 800M keys on OSM,
which has a single sharp drop and partial recovery before it declines again
(\S\ref{sec:scale-800m}).

\noindent\textbf{Oscillating.} ALEX repeatedly drops and recovers. Its throughput swings
between 2.05 and 3.45\,Mops/s within a single run (Figure~\ref{fig:sawtooth}). AlexOL
does the same under concurrency (\S\ref{sec:concurrent-scale}). The drops fall in the
batches the expansion schedule of \S\ref{sec:repair-loop} places on the key count.

\noindent\textbf{Mild.} BTree declines gently and evenly (retention 0.75 to 0.79 at
the default mix). ARTunsync is also smooth from batch to batch, but its retention varies
much more across datasets, from 1.19 on libio down to 0.65 on uniform.

The shape adds information that the size of the decline does not. Two indexes can lose
a similar fraction of throughput in very different ways.

\subsection{ALEX, LIPP and DILI Repair Themselves Differently}
\label{sec:repair-loop}

ALEX, LIPP and DILI can each repair their own structure while running. We describe
repair in three parts. A \textbf{signal} tells the index something has gone wrong, a
\textbf{repair action} is what it does in response, and a \textbf{limit} says how
often it may act. Table~\ref{tab:repair} gives the conditions we found in each of
those three implementations. SALI we report only by its throughput.

\begin{table}[t]
\small
\setlength{\tabcolsep}{3pt}
\caption{How ALEX, LIPP and DILI decide to repair themselves, taken from each source.}
\label{tab:repair}
\begin{tabular}{@{}l>{\raggedright\arraybackslash}p{2.6cm}>{\raggedright\arraybackslash}p{2.4cm}>{\raggedright\arraybackslash}p{1.5cm}@{}}
\toprule
\textbf{Index} & \textbf{Signal it watches} & \textbf{What it repairs} & \textbf{Limit} \\
\midrule
ALEX & node is 80\% full (or shifts per insert above 100, checked every 64 inserts) & expand the node in place, or refit or split it when its cost is 1.5$\times$ the prediction & none \\
DILI & lookups per key 2$\times$ worse than at last repair & that leaf only & \textbf{4 repairs, then never again} \\
LIPP & subtree doubled since it was built and 10\% of inserts since then hit a stored key & whole subtree, rebuilt from scratch & \textbf{subtrees built over 1M+ keys never rebuild} \\
\bottomrule
\end{tabular}
\vspace{-3mm}
\end{table}

\noindent\textbf{ALEX: the trigger is fill level, and it keeps firing.} ALEX repairs a
data node when the node is 80\% full. Nothing happens before that point, apart from a
check every 64 inserts for a runaway node whose inserts shift more than 100 keys.
Once the node is full, ALEX compares what its lookups actually cost against what the
node's model predicted when it was built. If the real cost is within 1.5$\times$ of the
prediction, it expands the node in place to 60\% full and stretches the model without
refitting it. If the cost is worse than that, or the node is at its maximum size, it
refits the model or splits the node. So fill level decides when a repair happens and
measured cost decides what kind. Nothing caps how many times this may happen. ALEX's
resize counter is kept only for statistics and is never read by any decision.

Because bulk loading fills every data node to 70\%, all of them reach 80\% after the
same relative growth. After a bulk load of $N_0$ keys the expansions therefore fire
together once the table holds $N_0 \cdot 8/7$ keys, and again at every further factor
of $4/3$. Our companion paper~\cite{namuganga2026synchronized} derives this schedule
from the three constants, confirms it by changing them, and shows that under
concurrency an operation that reaches a node being rebuilt restarts. The troughs in
ALEX's throughput in this paper fall in batches on that schedule
(\S\ref{sec:repair-cost}, \S\ref{sec:scale-800m}).

\noindent\textbf{LIPP skips a subtree built over more than a million keys.} LIPP rebuilds a subtree when
two things hold at once. The subtree must have doubled in size since it was built, and
at least 10\% of the inserts into it since then must have landed on a stored key, which
is what makes LIPP create a child node. But before either is checked, LIPP skips any
subtree that was built over more than a million keys. A node built that way is marked
fixed, and a fixed node does not trigger a rebuild. The bulk load builds the root over
the keys it is given, 50M in the default run, so that root is marked fixed.

\noindent\textbf{DILI stops repairing a leaf after four repairs.} DILI compares lookups per key now against
lookups per key at the last repair, and rebuilds the leaf when that doubles. After a
node has been repaired four times, DILI marks it permanently. Its repairs are also
limited to leaves.

On the runs, DILI retains 0.62 against LIPP's 0.56 on libio and 0.64 against 0.60 on
Planet, and it gains less from a rebuild (+58\% against +66\% on libio,
\S\ref{sec:reindex}). Both end below the BTree growth floor. LIPP retains 0.56 on
uniform, the same as on libio. ALEX stays below the BTree growth floor on every dataset
at the default 200M mix, and at 800M keys on books it ends lower than LIPP (retention
0.58 against 0.68, trend 0.62 against 0.69).

\subsection{Repair Is Not Free}
\label{sec:repair-cost}

\begin{figure*}[!t]
  \centering
  \includegraphics[width=\textwidth]{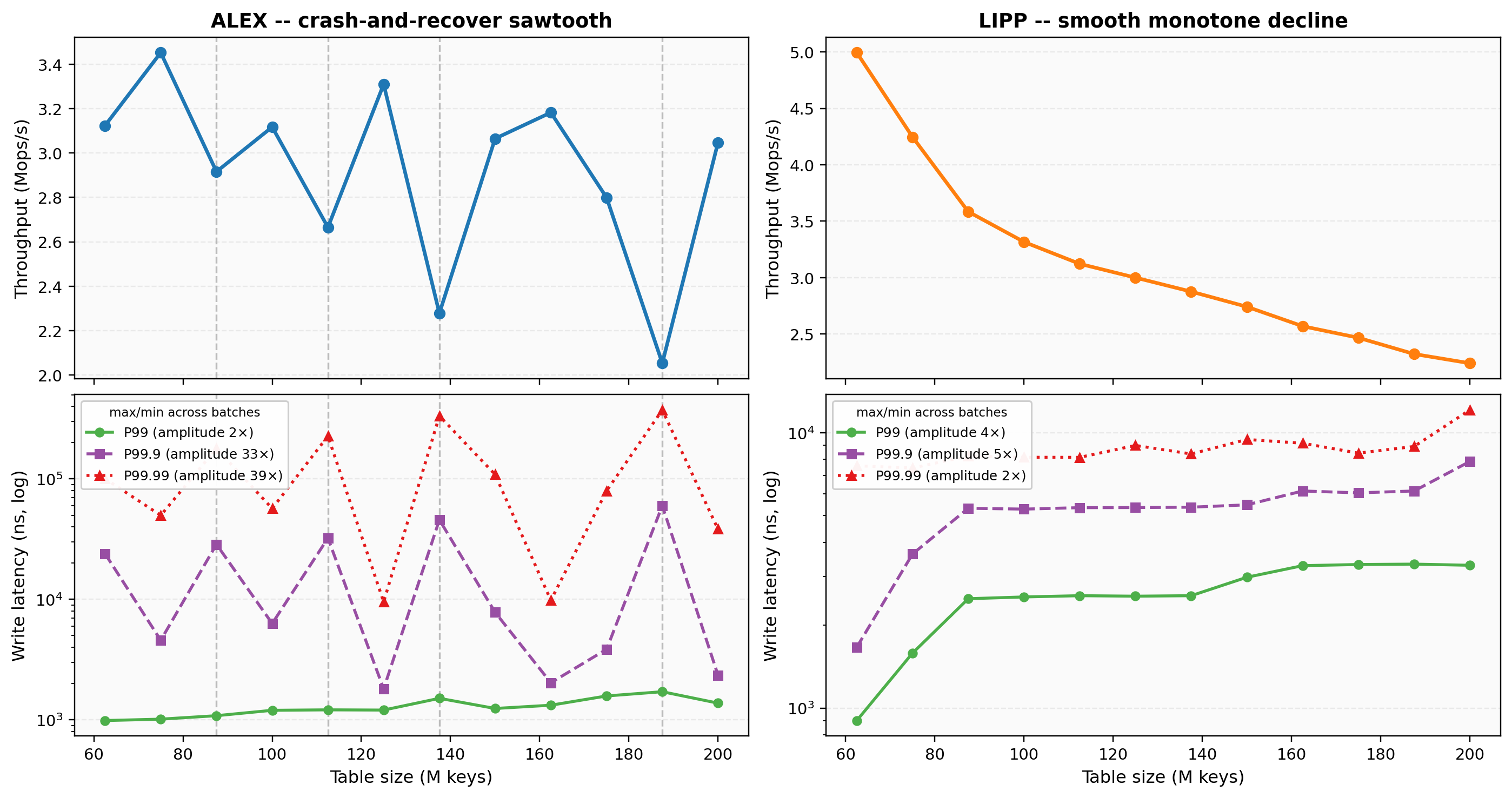}
  \Description{Four panels. Left column ALEX, right column LIPP. Top row per-batch
  throughput, bottom row per-batch write latency percentiles on a log scale. ALEX's
  write latency spikes where its throughput dips. LIPP's throughput declines
  smoothly while its write latency climbs steadily.}
  \vspace{-3mm}
  \caption{Per-batch throughput (top) and write tail latency (bottom) on \textbf{libio}
           at 25M ops/batch, mean across seeds 1866, 5, and 72. Dashed lines mark
           ALEX's throughput troughs, which fall in the batches where the expansion
           schedule of \S\ref{sec:repair-loop} predicts that the bulk-loaded nodes
           expand together. ALEX's write P99.9 spikes in those batches and
           swings 33$\times$ between its best and worst batch, while its read P99 barely
           moves (drift 1.21$\times$, not shown). LIPP's write P99.9 rises steadily with no
           spikes. For an oscillating series the ratio of last to first batch says
           little, so for ALEX we report the ratio of the largest to the smallest
           batch.}
  \label{fig:sawtooth}
\end{figure*}

Figure~\ref{fig:sawtooth} plots throughput and write latency per batch. The dips in
ALEX's throughput line up with spikes in its write latency (correlation $-0.88$ across
batches). P99.9 write latency swings 33$\times$ between its best and worst batch, from
1{,}808 to 59{,}909\,ns. Read latency barely moves (P99 drift 1.21$\times$). In the
trough batches P99.9 peaks at 29 to 60\,\textmu s, P99.99 reaches 180 to 370\,\textmu s,
and the maximum reaches several milliseconds. With 1\% of 12.5\,M inserts sampled,
P99.9 is the 125th largest sampled insert. Our companion paper, which times every
insert, shows that a percentile over a large batch misses the operations inside a
burst~\cite{namuganga2026synchronized}. The dashed lines in
Figure~\ref{fig:sawtooth} mark the troughs. They fall in the batches that contain
76.2, 101.6, 135.4 and 180.6\,M keys. For a bulk load of 50\,M keys the schedule puts
an expansion at 57.1\,M keys and then at those four counts~\cite{namuganga2026synchronized}.

LIPP is the opposite. Its write P99.9 climbs steadily, 4.7$\times$ over the run with no
spikes, and its \emph{read} latency ends up worse than ALEX's ever gets (P99.9 drift
5.5$\times$ against 1.22$\times$).

The two series move differently. ALEX keeps more of its throughput and its write tail
swings. LIPP's write tail climbs without spikes and its throughput keeps falling. An
operator with a write latency target can see both in the same figure.

This also shows why we report both retention and spread (\S\ref{sec:metrics}). On this
25M-ops/batch run ALEX's retention is 0.98, against 0.75 at the default 50M
(Table~\ref{tab:metrics}), while its spread is 0.123, the worst of any index.
Either number alone gives the wrong impression.

At our default settings ALEX retains 0.68 to 0.75, more than LIPP's 0.55 to 0.61, and
its write P99.9 swings 33 times between its best and worst batch. LIPP's read latency
ends up worse than ALEX's ever gets, 5.5 times against 1.22 times. Rebuilding the whole
index from outside costs 25 to 106 seconds at 200M keys (\S\ref{sec:reindex}), and most
of that is generic sorting.

Our companion paper tests two changes on AlexOL~\cite{namuganga2026synchronized}.
Bulk loading nodes to different fill levels leaves the same number of expansions, in
different batches, and under concurrency the count of inserts slower than
100\,\textmu s rises by 20 to 60\%. A percentile over a large batch reports that
spread-out run as a large improvement. Taking the repair off the inserting thread, with
a small buffer that exists only while a node is rebuilt and background threads that run
the rebuild, removes most of those slow inserts and keeps ALEX-OL's run time on the
400M books workload. XIndex~\cite{tang2020xindex} and FINEdex~\cite{li2021finedex} keep
a permanent buffer. On that same workload at sixteen threads they take 2.7 and 2.0
times as long as ALEX-OL, use more memory, and XIndex hung in four of nine runs while
FINEdex lost keys in two.

\subsection{Limitations}
\label{sec:limitations}

ARTunsync's retention on the synthetic uniform dataset is its lowest anywhere, and its
memory per key also grows most on that dataset. Periodic rebuilding recovers less on
Planet than on libio for every fitted index, and the Planet rebuilds cost more.

Our runs also have limits of scope. Each run is far shorter in wall-clock time than the
months a production store runs, so we measure the effect of many inserts rather than of
long elapsed time. The update and delete workloads cover only the indexes
whose update and delete paths we verified, and we report no scan workload. We evaluate
in-memory indexes only.
The default check on every run counts the success flags the wrappers return and
does not compare the payloads that lookups return. A later audit of the update
and delete paths, and a check against expected key state on the delete mix, is
what found the ART fault in \S\ref{sec:delete-inclusive}. Reading every inserted
key after a run, as our companion paper does, is a stronger test and is what
exposes losses that success flags miss~\cite{namuganga2026synchronized}.

\section{Discussion}
\label{sec:discussion}

\noindent\textbf{What is general and what is specific to learned indexes.}
The batched, never-reset execution model is entirely general. It is equally well defined
for BTree and ARTunsync, and we use it throughout as the baseline that says what table
growth alone costs. Finding~2 separates growth from the cost of inserting directly.
Runs that end with nearly the same table differ by 82\% for LIPP and 92\% for DILI depending on how many
keys arrived by insert, while ALEX moves 8\%, BTree 1\%, and ARTunsync 3\%.
The rebuild cost breakdown shows a related split, since the snapshot and sort every
rebuild pays is the same for every index, 83 to 84\% of BTree's rebuild cost against 19
to 26\% of ALEX's.
So the way we \emph{measure} degradation is general-purpose, while the large
insert-driven losses we see come from LIPP and DILI, not from every learned index.

\noindent\textbf{What to report.}
Learned-index work targeting storage should report \textbf{retention}, \textbf{trend},
\textbf{spread} and \textbf{P99 latency drift} alongside peak throughput.
Peak throughput describes an index at $t{=}0$ under conditions no deployed system stays
in, and learned structures are proposed as memtable and block-index replacements in
systems such as RocksDB~\cite{dong2021rocksdb} that run for months.
The four together cost one longitudinal run to produce and are what separates an index
that settles from one that does not.

\noindent\textbf{What to do with a single-shot ranking.}
Treat it as a snapshot of $t{=}0$ rather than a prediction of $t{=}N$.
If the workload is insert-heavy and long-running, check retention and trend before
trusting a reported lead, because a large one can and does invert.
If a candidate degrades, check which \emph{shape} it has, since a smooth steady decline
is a different operational risk from a drop and recover sawtooth at the same retention,
and the second shows up as write latency spikes rather than as a falling average.
If periodic rebuilding is on the table, price it rather than assuming it is free. In
our runs it recovers most of the loss for LIPP and DILI, and it costs 25 to 106 seconds
at 200M keys, much of it sorting that any index would pay.
None of this requires abandoning single-shot benchmarks. It requires not treating their
output as the whole answer for any index whose state accumulates.

\section{Conclusion}
\label{sec:conclusion}

Learned indexes are usually benchmarked at their best moment, just after they are
built. We ran nine indexes across five datasets without ever resetting them, and found
that this moment does not last.
Nine of the twelve single-threaded learned index and dataset pairs lead ARTunsync at the
first batch, and five of those are behind it by the last, so an index chosen on a
single-shot number is not always the index that performs best later.

The decline is not an artifact of one setting. LIPP, and DILI where we ran it, lose more
than the traditional baselines at read-heavy and write-heavy mixes, across a fourfold
range of batch sizes, and with updates and deletes. SALI loses more than
ARTolc at every thread count from 8 to 32. Out to 800M keys and 24 batches, LIPP's
worst throughput still falls at the last batch.
For LIPP and DILI the decline is caused mainly by inserting rather than by holding
more keys. Runs that end with nearly the same table differ by 82\% and 92\% depending
on how many keys arrived by insert, while ALEX differs by 8\% and ARTunsync and BTree
differ by 1 to 3\%.
And it is recoverable at a price. Rebuilding every other batch recovers most of what
LIPP and DILI lost, for 25 to 106 seconds per rebuild at 200M keys.
On the write tail, ALEX swings and LIPP climbs. Our companion paper finds that
spreading ALEX's repairs out does not remove their cost, and that taking them off the
inserting thread removes most of the latency spikes on its 400M books
workload~\cite{namuganga2026synchronized}.


\section*{Use of AI Tools}
We used AI assistants, Anthropic's Claude and OpenAI's Codex, to edit the text, to write
and run analysis and experiment scripts, and to check this paper against our companion
paper. The authors reviewed this work and take full responsibility for the contents of
the paper.

\clearpage
\bibliographystyle{ACM-Reference-Format}
\bibliography{main_arxiv}

\appendix

\section{Additional Results}

This appendix holds the per-experiment detail behind Table~\ref{tab:robustness}.
Every result uses the protocol of \S\ref{sec:method} with three seeds (1866, 5, 72).

\subsection{Degradation Trajectories}
\label{app:trajectories}

\begin{figure*}[!t]
  \centering
  \includegraphics[width=\textwidth]{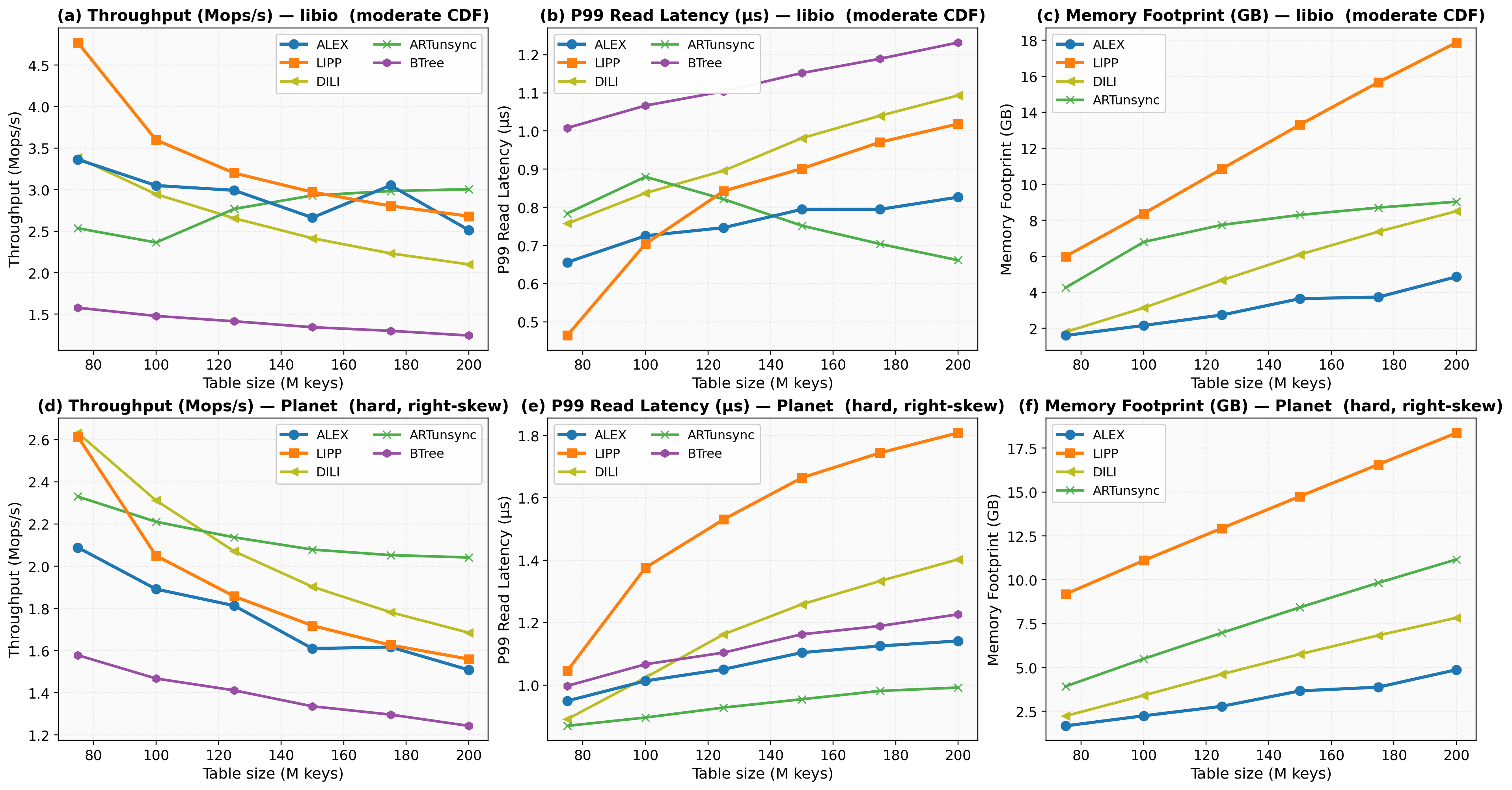}
  \Description{Three-panel figure showing per-batch throughput, P99 read latency, and index-reported memory footprint on libio.}
  \vspace{-4mm}
  \caption{The three per-batch time-series the longitudinal benchmark produces, on libio
           (top row, moderate CDF) and Planet (bottom row, hard right-skew): (a, d)
           throughput, (b, e) P99 read latency, (c, f) memory footprint. Mean across 3
           seeds (1866, 5, 72). On libio LIPP loses 44\% of its throughput (4.77 to
           2.68\,Mops/s), its P99 read latency drifts 2.2$\times$ and its memory grows
           3.0$\times$ (6.0 to 17.9\,GB). On Planet every learned index ends between BTree
           and ARTunsync. Memory is each index's self-reported structural size, and BTree
           does not report it.}
  \label{fig:gallery}
\end{figure*}

\begin{table*}[t]
\small
\setlength{\tabcolsep}{3.2pt}
\caption{Longitudinal statistics for the single-threaded indexes on all five datasets,
  mean across seeds 1866, 5 and 72. \textbf{Ret} is throughput retention (last batch over
  first), \textbf{Trd} is trend (mean of the last two batches over the mean of the first
  two), \textbf{Spr} is spread (residual coefficient of variation around a Theil--Sen
  trend line) and \textbf{P99} is read tail-latency drift (last batch over first). DILI
  was run on libio and Planet only.
  BTree, set apart below the rule, is the growth floor, meaning what an index with no fitted
  model pays for the table getting bigger. \textbf{Bold} marks a retention below BTree
  on the same dataset. Every
  learned index is bold on every dataset it appears on. ARTunsync is bold only on uniform, where it
  posts its worst retention anywhere. Its Ret above 1 on libio means it got faster over
  the run. This table is also an example of the reporting we recommend in
  \S\ref{sec:discussion}.}
\label{tab:metrics}
\begin{tabular}{@{}l rrrr rrrr rrrr rrrr rrrr@{}}
\toprule
 & \multicolumn{4}{c}{\textbf{libio}} & \multicolumn{4}{c}{\textbf{Planet}}
 & \multicolumn{4}{c}{\textbf{COVID}} & \multicolumn{4}{c}{\textbf{OSM}}
 & \multicolumn{4}{c}{\textbf{uniform}} \\
\cmidrule(lr){2-5}\cmidrule(lr){6-9}\cmidrule(lr){10-13}\cmidrule(lr){14-17}\cmidrule(lr){18-21}
\textbf{Index} & Ret & Trd & Spr & P99 & Ret & Trd & Spr & P99 & Ret & Trd & Spr & P99
 & Ret & Trd & Spr & P99 & Ret & Trd & Spr & P99 \\
\midrule
ALEX      & \textbf{0.75} & 0.87 & 0.06 & 1.3 & \textbf{0.72} & 0.79 & 0.03 & 1.2 & \textbf{0.73} & 0.84 & 0.06 & 1.3 & \textbf{0.73} & 0.78 & 0.02 & 1.5 & \textbf{0.68} & 0.85 & 0.09 & 1.2 \\
LIPP      & \textbf{0.56} & 0.66 & 0.11 & 2.2 & \textbf{0.60} & 0.68 & 0.09 & 1.7 & \textbf{0.55} & 0.62 & 0.06 & 1.7 & \textbf{0.61} & 0.73 & 0.06 & 1.4 & \textbf{0.56} & 0.62 & 0.07 & 1.6 \\
DILI      & \textbf{0.62} & 0.68 & 0.03 & 1.4 & \textbf{0.64} & 0.70 & 0.04 & 1.6 & --   & --   & --   & --  & --   & --   & --   & --  & --   & --   & --   & --  \\
ARTunsync & 1.19 & 1.22 & 0.04 & 0.8 & 0.88 & 0.90 & 0.02 & 1.1 & 0.79 & 0.83 & 0.03 & 1.2 & 0.77 & 0.82 & 0.01 & 1.2 & \textbf{0.65} & 0.68 & 0.05 & 1.3 \\
\midrule
BTree     & 0.79 & 0.83 & 0.01 & 1.2 & 0.79 & 0.83 & 0.02 & 1.2 & 0.76 & 0.82 & 0.01 & 1.2 & 0.75 & 0.81 & 0.01 & 1.2 & 0.77 & 0.82 & 0.01 & 1.2 \\
\bottomrule
\end{tabular}
\vspace{-3mm}
\end{table*}

Figure~\ref{fig:gallery} shows the three time-series the longitudinal protocol produces
per index across two datasets.
The metrics themselves are the standard ones every benchmark reports.
What is new is seeing each of them per batch over a run that is never reset, so we see
the trajectory and not just one value.

\textbf{Throughput (panels a, d).}
On libio, LIPP drops 44\%, from 4.77 to 2.68\,Mops/s, which turns an 88\% initial
advantage over ARTunsync into an 11\% deficit by batch~6.
On Planet every learned index ends between BTree (1.24\,Mops/s) and ARTunsync (2.04),
and the first-batch leads of LIPP and DILI over ARTunsync are gone.

\textbf{Tail latency (panels b, e).}
On libio, LIPP's P99 read latency drifts 2.2$\times$ (0.46 to 1.02\,\textmu s) and DILI's
1.4$\times$. On Planet, LIPP drifts 1.7$\times$ and DILI 1.6$\times$. ARTunsync is the
only index whose read tail improves over the run on libio (0.8$\times$).

\textbf{Memory (panels c, f).}
Index-reported memory grows for every structure, but the table itself grows $2.7\times$
across the window, so cost \emph{per key} is the better measure.
On libio, LIPP rises from 80.0 to 89.4 bytes/key (+12\%) and DILI from 23.9 to 42.5
(+78\%), so their structures grow faster than the number of keys.
ARTunsync instead \emph{falls} from 56.8 to 45.2 bytes/key as the denser trie shares more
prefixes.
On Planet, DILI again rises, from 29.8 to 39.2 (+32\%), while LIPP starts at a very high
122 bytes/key and falls toward 92.
Absolute footprints are not comparable across families (ART holds pointers to pairs stored
outside the index, while learned indexes store them internally), but within-index trends
are.

\begin{table*}[t]
\small
\setlength{\tabcolsep}{4pt}
\caption{Whether a concurrent learned index that leads at the first batch still leads at
  the last, measured against ARTolc on four datasets and four thread counts (mean across
  seeds 1866, 5 and 72). Each cell gives the margin at the first batch and at the last,
  so $+$20 and $-$38 means the index opened 20\% faster than ARTolc and closed 38\%
  slower. \textbf{Bold} marks a reversal, a lead at the first batch that is gone by the
  last. There are nine leads in the table and seven of them reverse. AlexOL on COVID at
  32 threads opens level with ARTolc ($+$0.1\%) and ends 18\% behind, and we count it as
  neither a lead nor a reversal. Planet and OSM contribute no leads because ARTolc is
  already fastest at the first batch there, the same pattern those two datasets show
  single-threaded (Table~\ref{tab:reversal}).}
\label{tab:concurrent}
\begin{tabular}{@{}ll rr rr rr rr@{}}
\toprule
 & & \multicolumn{2}{c}{\textbf{8 threads}} & \multicolumn{2}{c}{\textbf{16 threads}}
 & \multicolumn{2}{c}{\textbf{24 threads}} & \multicolumn{2}{c}{\textbf{32 threads}} \\
\cmidrule(lr){3-4}\cmidrule(lr){5-6}\cmidrule(lr){7-8}\cmidrule(lr){9-10}
\textbf{Dataset} & \textbf{Index} & First & Last & First & Last & First & Last & First & Last \\
\midrule
libio   & AlexOL & \textbf{$+$17} & \textbf{$-$17} & \textbf{$+$7} & \textbf{$-$20} & $-$0 & $-$33 & $-$1 & $-$43 \\
        & SALI   & \textbf{$+$20} & \textbf{$-$38} & \textbf{$+$9} & \textbf{$-$36} & \textbf{$+$12} & \textbf{$-$40} & \textbf{$+$13} & \textbf{$-$41} \\
\addlinespace
COVID   & AlexOL & $+$21 & $+$9 & $+$7 & $+$2 & \textbf{$+$3} & \textbf{$-$9} & $+$0 & $-$18 \\
        & SALI   & $-$0 & $-$26 & $-$12 & $-$27 & $-$4 & $-$27 & $-$4 & $-$27 \\
\addlinespace
Planet  & AlexOL & $-$11 & $-$26 & $-$21 & $-$25 & $-$25 & $-$37 & $-$28 & $-$45 \\
        & SALI   & $-$28 & $-$45 & $-$37 & $-$40 & $-$29 & $-$46 & $-$36 & $-$46 \\
\addlinespace
OSM     & AlexOL & $-$27 & $-$27 & $-$27 & $-$32 & $-$35 & $-$37 & $-$35 & $-$43 \\
        & SALI   & $-$41 & $-$46 & $-$33 & $-$49 & $-$40 & $-$48 & $-$36 & $-$49 \\
\bottomrule
\end{tabular}
\vspace{-3mm}
\end{table*}

\subsection{Longitudinal Statistics on All Five Datasets}
\label{app:stats}
\label{sec:stats}

Table~\ref{tab:metrics} condenses the trajectories into the four summary statistics for
every single-threaded index on all five datasets. Reporting all five datasets at once is
what lets the next three observations speak about the indexes rather than only about
libio and Planet.

\textbf{BTree sets a growth floor, and it barely moves.}
With no model to go stale, BTree retains 0.75--0.79 and drifts $1.2\times$ at the read
tail at the default mix, synthetic and real alike.
That narrow band is what table growth alone costs an index, and it is the line the other
indexes should be read against.

\textbf{LIPP sits below that floor everywhere.}
Its retention is 0.55 on COVID, 0.56 on libio and uniform, 0.60 on Planet and 0.61 on
OSM, a six-point band that never reaches BTree's worst dataset.
DILI, run on libio and Planet only, is also below the floor at 0.62 and 0.64.
ALEX is below it too, but closer, at 0.68--0.75.
The excess below the growth floor, not the decline itself, is the honeymoon cost, and
LIPP pays it on the synthetic uniform dataset as much as on a hard real one.

\textbf{ARTunsync varies the most across datasets.}
Its retention runs from 1.19 on libio down to 0.65 on uniform, a far wider range than
any learned index shows.
Its memory per key also grows most on uniform (+52\%, against $-$20\% to +7\%
elsewhere).

Trend moves with retention on every dataset, which suggests the retention values are not
driven by one unusual endpoint, and in one cell the difference matters. ALEX on uniform
retains 0.68 but trends at 0.85, because its last batch drops sharply (from 2.84 to
2.03\,Mops/s) after a rise. Retention alone would call that its worst dataset, but read
with trend it is not. That is the case the pair of columns exists for.
Spread separates the two learned indexes from each other rather than from the
baselines: LIPP swings at 0.06--0.11 while DILI sits at 0.03--0.04, as steady as BTree
and ARTunsync even though it degrades further than they do.

\subsection{Concurrent Indexes}
\label{app:concurrency}
\label{sec:concurrent}

Table~\ref{tab:concurrent} repeats the reversal test of \S\ref{sec:honeymoon} for the
concurrent indexes, on four datasets at 8, 16, 24 and 32 threads.
There are nine cases where a learned index leads at the first batch and seven of them are
behind by the last, plus one case that opens level and ends behind, so the reversal is
not an artifact of single-threaded execution.

\textbf{Concurrency moves the crossing point.}
On libio SALI opens ahead of ARTolc at all four thread counts and finishes behind every
time, so there the reversal is present from the lowest concurrency we test.
COVID shows the opposite dependence and is the more informative case.
AlexOL's first-batch lead there shrinks monotonically as threads are added, from
$+$21\% at 8 threads to $+$7\%, $+$3\% and $+$0\%, and it reverses only at 24 and 32
threads while keeping its lead at 8 and 16.
So added threads do not uniformly help or hurt the learned index. They help ARTolc more,
and the crossing arrives earlier in the run as concurrency rises. On libio that is
already enough to invert the ranking at 8 threads, and on COVID it takes 24.
Planet and OSM contribute no cases at all, because ARTolc is fastest from the first
batch at every thread count, exactly as those two datasets behave single-threaded
(Table~\ref{tab:reversal}).

\textbf{The learned indexes degrade at every thread count, while ARTolc holds up
better as threads are added.} SALI retains 0.58--0.68 on libio and 0.59--0.74 on COVID
across 8--32 threads, and AlexOL 0.74--0.85 and 0.72--0.80, so their decline barely
moves with concurrency.
ARTolc's retention is higher at 32 threads than at 8 on every dataset, from 1.12 to
1.30 on libio, 0.80 to 0.97 on COVID, 0.87 to 0.95 on Planet and 0.75 to 0.87 on OSM,
although not always rising at every step in between. Only on libio does it go above 1.
This is what moves the crossings above earlier. BTrOLC degrades at every thread count
too, with retention between 0.75 and 0.84.
Adding threads raises every curve, but it does not change the downward trajectory of the
learned indexes, so more threads do not offset the decline.
That holding up is throughput. It is not a clean tail. Our companion paper times every
insert on a 400M-key books run at sixteen threads and finds ART-OLC has the most
inserts slower than 100\,\textmu s of any index, 103{,}187 of them, spread across every
batch rather than bunched in bursts, and that its worst-to-median write p99.9 grows
from $2.2\times$ at one thread to $61.4\times$ at sixteen~\cite{namuganga2026synchronized}.
So ARTolc is the baseline that keeps its mean rate as threads are added, and it is also
the baseline that pays the most high-latency inserts.

\subsection{Concurrency at Scale: 800M Keys, 8 Threads}
\label{sec:concurrent-scale}

\begin{figure*}[!t]
  \centering
  \includegraphics[width=\textwidth]{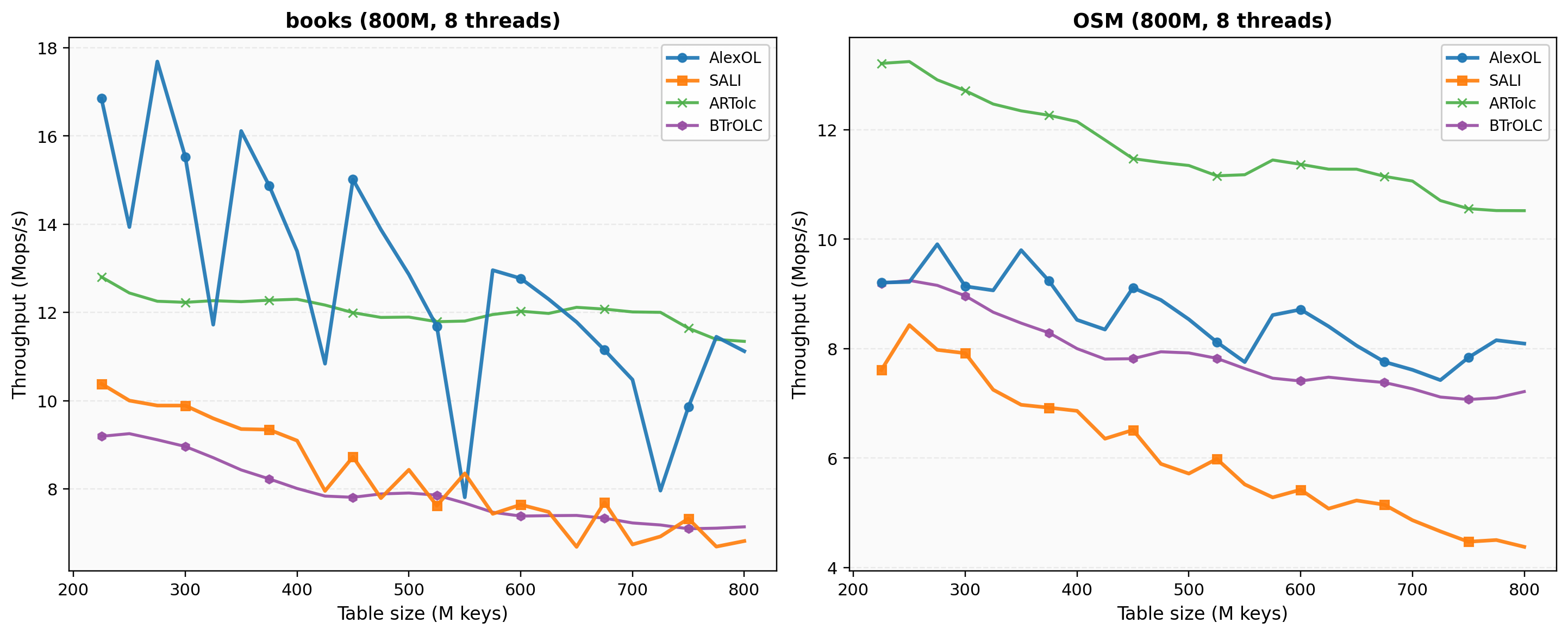}
  \Description{Two line charts showing mean throughput across three seeds vs table size
  from 200M to 800M keys, for AlexOL, SALI, ARTolc, and BTrOLC at 8 threads on books and
  OSM.}
  \vspace{-3mm}
  \caption{Per-batch throughput from 200M to 800M keys (24 batches, 8 threads), mean
           across three seeds (1866, 5, 72), on \textbf{books} (left)
           and \textbf{OSM} (right).
           AlexOL shows a drop and recover sawtooth on books, of similar size to
           single-threaded ALEX's (\S\ref{sec:scale-800m}).
           SALI retains 0.66 on books, level with AlexOL, and 0.58 on OSM, the lowest of
           the four there.
           ARTolc, whose retention was above 1 on libio at 200M (up to 1.30 at 32
           threads), declines at this scale (retention 0.89 on books, 0.80 on OSM), the
           same pattern as ARTunsync and BTree in \S\ref{sec:scale-800m}.
           On books AlexOL and ARTolc finish about level. That ranking is for this
           balanced 8-thread run, not for the companion paper's insert-only 16-thread
           run at 400M keys.}
  \label{fig:scale-concurrent}
\end{figure*}

Figure~\ref{fig:scale-concurrent} runs the concurrent indexes on the same 800M-key,
24-batch protocol used for the single-threaded scale extension (\S\ref{sec:scale-800m}),
to check whether the concurrency and scale results hold at the same time.
They do. Every index declines from the first batch to the last, and ARTolc's
retention above 1 on libio at 200M (\S\ref{sec:concurrent}) does not carry over to
800M, where it declines like every other index. On books AlexOL finishes about
level with ARTolc. That is a balanced mix at 8 threads, on the released binary, out
to 800M keys. It is not the same race as the companion paper's 400M books run,
which is insert-only at 16 threads and uses the patched ALEX-OL. There ALEX-OL
finishes the inserts in 17.2\,s against ART-OLC's 21.8\,s and has the lowest
read-heavy lookup latency of the concurrent
indexes~\cite{namuganga2026synchronized}. Who looks faster depends on the mix, the
thread count, the scale, and which ALEX-OL binary is running.

\subsection{Workload Spectrum}
\label{app:spectrum}
\label{sec:workload-spectrum}

\begin{figure*}[!t]
  \centering
  \includegraphics[width=\textwidth]{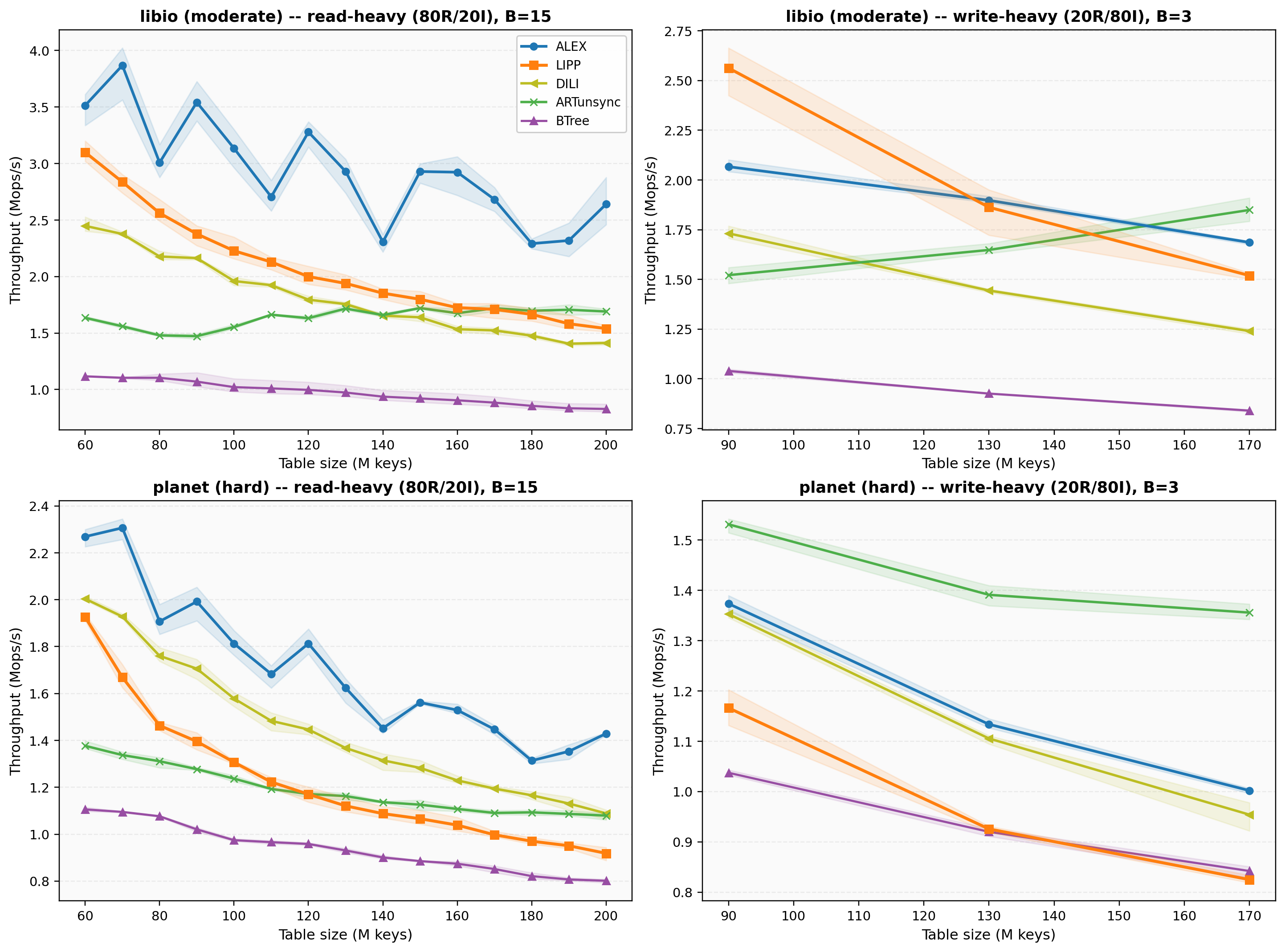}
  \Description{Four line charts showing per-batch throughput for ALEX, LIPP, DILI,
  ARTunsync, and BTree under a read-heavy (80\% read / 20\% insert) and a write-heavy
  (20\% read / 80\% insert) mix, on libio and planet, mean across three seeds.}
  \vspace{-3mm}
  \caption{Per-batch throughput under \textbf{read-heavy} ($r_\text{read}{=}0.8$,
           $B{=}15$, left) and \textbf{write-heavy} ($r_\text{read}{=}0.2$, $B{=}3$,
           right) mixes, on \textbf{libio} (top) and \textbf{planet} (bottom). Shaded
           bands are the range across seeds 1866, 5, and 72. Both mixes still bulk-load
           50M and grow to 200M keys, so they differ from the balanced main result
           (\S\ref{sec:honeymoon}, $B{=}6$) in the read and insert mix and in the number
           of batches.
           LIPP's reversal against ARTunsync appears at \emph{both} ends of the
           spectrum. On libio it goes from a 90\% first-batch lead to a 9\% deficit
           (read-heavy) and, in just 3 batches, from a 68\% lead to an 18\% deficit
           (write-heavy).
           On planet write-heavy LIPP never leads. It starts 24\% behind ARTunsync and
           ends 39\% behind.}
  \label{fig:workload-spectrum}
\end{figure*}

The evaluation so far uses a single point on the read/insert spectrum
($r_\text{read}{=}r_\text{insert}{=}0.5$). Following GRE's own workload-generality
sweep, we hold the longitudinal protocol fixed and rerun it at read-heavy
($r_\text{read}{=}0.8$, $r_\text{insert}{=}0.2$) and write-heavy
($r_\text{read}{=}0.2$, $r_\text{insert}{=}0.8$) on libio and planet, all five
single-threaded indexes, three seeds each. Both mixes still bulk-load 50M keys and
grow to 200M, so the batch count changes along with the mix. It is $B{=}15$ for
read-heavy (10M inserts per batch) and $B{=}3$ for write-heavy (40M inserts per batch),
against the balanced default's $B{=}6$ (25M inserts per batch).

\textbf{The ranking reversal is not specific to the balanced workload, and it can happen
in as few as three batches.} Figure~\ref{fig:workload-spectrum} shows LIPP ahead of
ARTunsync at the first batch in three of the four mix and dataset combinations, and
behind it by the last batch in all three.
On libio the read-heavy reversal goes from a 90\% lead to a 9\% deficit, and the
write-heavy one completes in three batches, from a 68\% lead to an 18\% deficit.
On that same libio read-heavy run ALEX starts at 3.51\,Mops/s against 1.63 for
ARTunsync and 1.12 for BTree, and by 200M keys ALEX has lost 25\% of its first-batch
throughput, LIPP 50\%, and DILI 42\%, while ARTunsync has gained 3\%. Those are the
read-heavy numbers. The paper's default balanced mix on the same dataset is the
one in Table~\ref{tab:metrics}, where ALEX loses 25\%, LIPP 44\%, and DILI 38\%.
On planet write-heavy LIPP is already 24\% behind ARTunsync at the first batch, so on
this combination the bulk-load peak is not enough to put the learned index ahead even
briefly.

\textbf{Retention is lower when the same inserts are spread over more batches.} All
three mixes insert the same 150M keys and grow the table from 50M to 200M, yet LIPP's
libio retention rises as there are fewer, larger batches, from 0.497 (read-heavy,
$B{=}15$) to 0.562 (balanced, $B{=}6$) to 0.593 (write-heavy, $B{=}3$). DILI shows the
same ordering (0.577, 0.621, 0.716). BTree changes less over the same sweep (0.741,
0.788, 0.808 on libio, a 9\% relative range against 19\% for LIPP and 24\% for DILI).
The mix and the batch count change together here, so this sweep alone cannot say which
of them matters. The batch-size sweep (\S\ref{sec:batchsize}) holds the mix fixed and
shows the same direction.

\textbf{On the hard dataset under heavy writes the traditional index comes out on
top.} Under planet write-heavy ARTunsync finishes as the fastest index
(1.36\,Mops/s), ahead of every learned index including ALEX (1.00).

\begin{table}[t]
\small
\setlength{\tabcolsep}{3pt}
\caption{Throughput retention (last/first batch) under read-heavy (RH) and write-heavy
  (WH) mixes, across all five datasets, mean across seeds 1866, 5, and 72. DILI is
  libio/planet-only for this axis (\S\ref{sec:workload-spectrum-datasets}) and omitted
  here.}
\label{tab:spectrum-datasets}
\begin{tabular}{@{}l rrrr rrrr@{}}
\toprule
 & \multicolumn{4}{c}{\textbf{Read-heavy}} & \multicolumn{4}{c}{\textbf{Write-heavy}} \\
\cmidrule(lr){2-5}\cmidrule(lr){6-9}
\textbf{Dataset} & ALEX & LIPP & ART & BTree & ALEX & LIPP & ART & BTree \\
\midrule
libio   & 0.75 & 0.50 & 1.03 & 0.74 & 0.82 & 0.59 & 1.22 & 0.81 \\
planet  & 0.63 & 0.48 & 0.78 & 0.73 & 0.73 & 0.71 & 0.89 & 0.81 \\
covid   & 0.90 & 0.49 & 0.70 & 0.71 & 0.91 & 0.66 & 0.87 & 0.86 \\
osm     & 0.70 & 0.66 & 0.72 & 0.69 & 0.82 & 0.76 & 0.89 & 0.86 \\
uniform & 1.11 & 0.54 & 0.62 & 0.76 & 0.91 & 0.59 & 0.72 & 0.86 \\
\bottomrule
\end{tabular}
\vspace{-3mm}
\end{table}

\subsection{Cross-Dataset Coverage and a Uniform Control}
\label{sec:workload-spectrum-datasets}

Table~\ref{tab:spectrum-datasets} extends the workload spectrum to the two remaining
real-world datasets (covid, osm) and to the synthetic \textbf{uniform} dataset, whose
keys are drawn uniformly at random and so have a flat CDF.
\textbf{Two libio and planet patterns repeat.}
BTree's retention stays in a narrow band regardless of dataset or mix (0.69--0.76
read-heavy, 0.81--0.86 write-heavy), the same tight growth floor seen at the balanced
mix (\S\ref{sec:stats}). And LIPP degrades more than both traditional baselines on all
five datasets at both mixes. The gap is smallest on OSM read-heavy, where LIPP retains
0.66 against 0.69 for BTree and 0.72 for ARTunsync, while on libio, planet and covid
its read-heavy retention is 0.48--0.50.

\textbf{The uniform dataset shows the effect does not need an irregular key
distribution.} A flat CDF is the easiest case for a model to fit, yet LIPP's uniform
retention (0.54 read-heavy, 0.59 write-heavy) is inside the same range as the
real-world datasets, not above it.
ALEX on uniform read-heavy retains 1.11, one of only two growth or delete runs where
ALEX's retention is above 1 (the other is the delete mix on libio,
\S\ref{sec:delete-inclusive}). At the balanced mix on uniform it retains 0.68, so we do
not read this as a general property of uniform data.

\subsection{Batch-Size Sensitivity}
\label{app:batchsize}
\label{sec:batchsize}

\begin{table}[t]
\small
\setlength{\tabcolsep}{4pt}
\caption{Retention and trend on \textbf{libio} as batch granularity varies, holding total
  inserted keys (150M) and every other protocol parameter fixed. Mean across seeds 1866,
  5, and 72. Trend is not reported at $B{=}3$, because with three batches the first two
  and the last two share the middle batch.}
\label{tab:batchsize}
\begin{tabular}{@{}l rr rr rr@{}}
\toprule
 & \multicolumn{2}{c}{\textbf{25M/batch}} & \multicolumn{2}{c}{\textbf{50M/batch}} & \multicolumn{2}{c}{\textbf{100M/batch}} \\
 & \multicolumn{2}{c}{($B{=}12$)} & \multicolumn{2}{c}{($B{=}6$, default)} & \multicolumn{2}{c}{($B{=}3$)} \\
\cmidrule(lr){2-3}\cmidrule(lr){4-5}\cmidrule(lr){6-7}
\textbf{Index} & Ret & Trd & Ret & Trd & Ret & Trd \\
\midrule
ALEX      & 0.98 & 0.78 & 0.75 & 0.87 & 0.86 & -- \\
LIPP      & 0.45 & 0.49 & 0.56 & 0.66 & 0.60 & -- \\
DILI      & 0.60 & 0.63 & 0.62 & 0.68 & 0.70 & -- \\
ARTunsync & 1.10 & 1.18 & 1.19 & 1.22 & 1.21 & -- \\
BTree     & 0.79 & 0.79 & 0.79 & 0.83 & 0.83 & -- \\
\bottomrule
\end{tabular}
\vspace{-3mm}
\end{table}

Does the result depend on the choice of 50M operations per batch? We answer this with
a sweep on libio in which total inserted keys (150M), the bulk-load fraction and the
read and insert mix are held fixed while operations per batch vary over
$M \in \{25\text{M}, 50\text{M}, 100\text{M}\}$ ($B \in \{12, 6, 3\}$), a 4$\times$ range.
Unlike the workload spectrum sweep (\S\ref{sec:workload-spectrum}), which changes the
batch count along with the mix, this sweep changes only the batch size.

\begin{figure*}[!t]
  \centering
  \includegraphics[width=\textwidth]{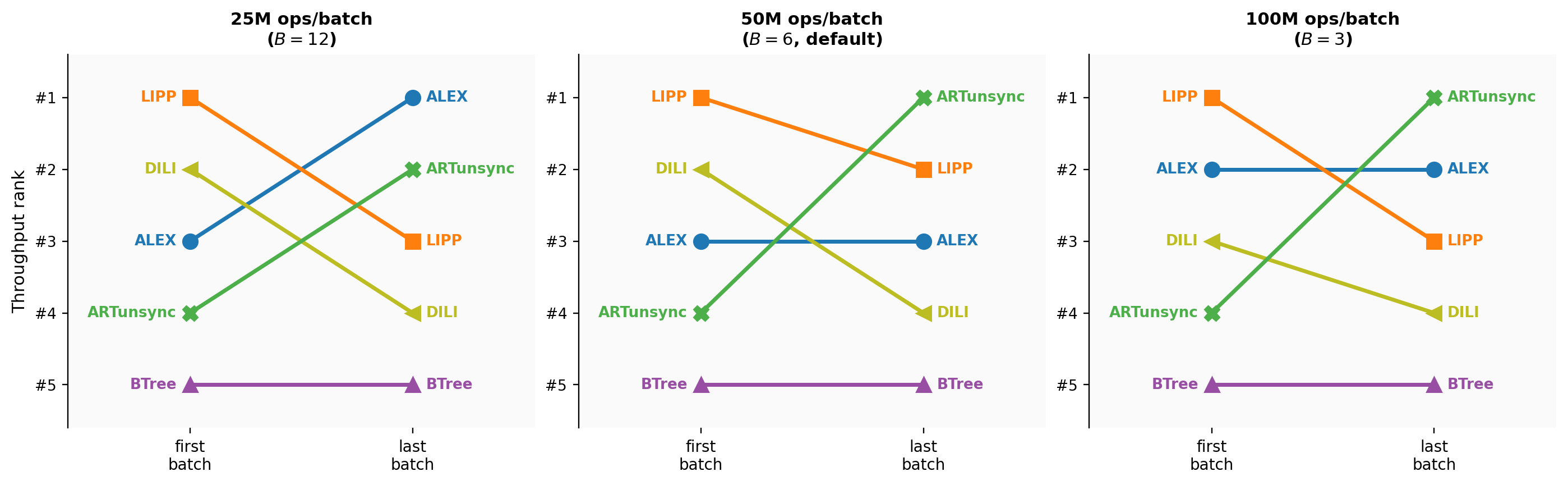}
  \Description{Three slope charts showing each index's throughput rank at the first
  batch and at the last batch, for 25M, 50M, and 100M operations per batch on libio.
  LIPP starts first and finishes lower in all three. ARTunsync starts fourth and
  finishes first or second in all three. BTree stays fifth throughout.}
  \vspace{-3mm}
  \caption{Throughput rank at the first batch vs.\ the last batch on \textbf{libio}, at
           three batch granularities: 25M ops/batch ($B{=}12$), 50M ($B{=}6$, the
           paper's default), and 100M ($B{=}3$). Mean across seeds 1866, 5, and 72.
           The crossing pattern is the same in all three panels. LIPP starts first and
           never finishes first, ARTunsync starts fourth and finishes first or second,
           DILI always falls, and BTree stays fifth.}
  \label{fig:batchsize}
\end{figure*}

\textbf{The ranking reversal does not depend on batch size.}
Figure~\ref{fig:batchsize} shows each index's rank at the first batch against its rank
at the last batch, once per batch size. LIPP is first at the first batch at all three
sizes and first at the last batch at none of them (89\%, 88\% and 66\% ahead of
ARTunsync at the first batch, and 23\%, 11\% and 18\% behind it at the last).
ARTunsync is fourth at the first batch every time and first or second at the last.
DILI always falls and BTree stays fifth. Across this range of batch sizes, the index
that looks best at the start is not the one that is best at the end.

\textbf{Retention shifts, but not enough to regroup the indexes.} Batch size does move
the retention numbers in Table~\ref{tab:batchsize}, most for LIPP (0.45--0.60 across
the three sizes) and least for BTree (0.79--0.83). But LIPP and DILI stay below BTree's
growth floor at every batch size. Retention is lower when there are more batches, the
same direction as in the workload spectrum.

\textbf{ALEX's retention is the least stable, because of its oscillation.} It is 0.98
at $B{=}12$, 0.75 at $B{=}6$ and 0.86 at $B{=}3$. Because ALEX's throughput oscillates
within a run, a statistic computed from the first and last batch alone depends on where
those two batches fall in the oscillation. For ALEX, retention should be read alongside
trend and spread rather than on its own.

\subsection{Bulk-Load Fraction and Path Dependence}
\label{app:rho}
\label{sec:rho-init}

\begin{figure*}[!t]
  \centering
  \includegraphics[width=\textwidth]{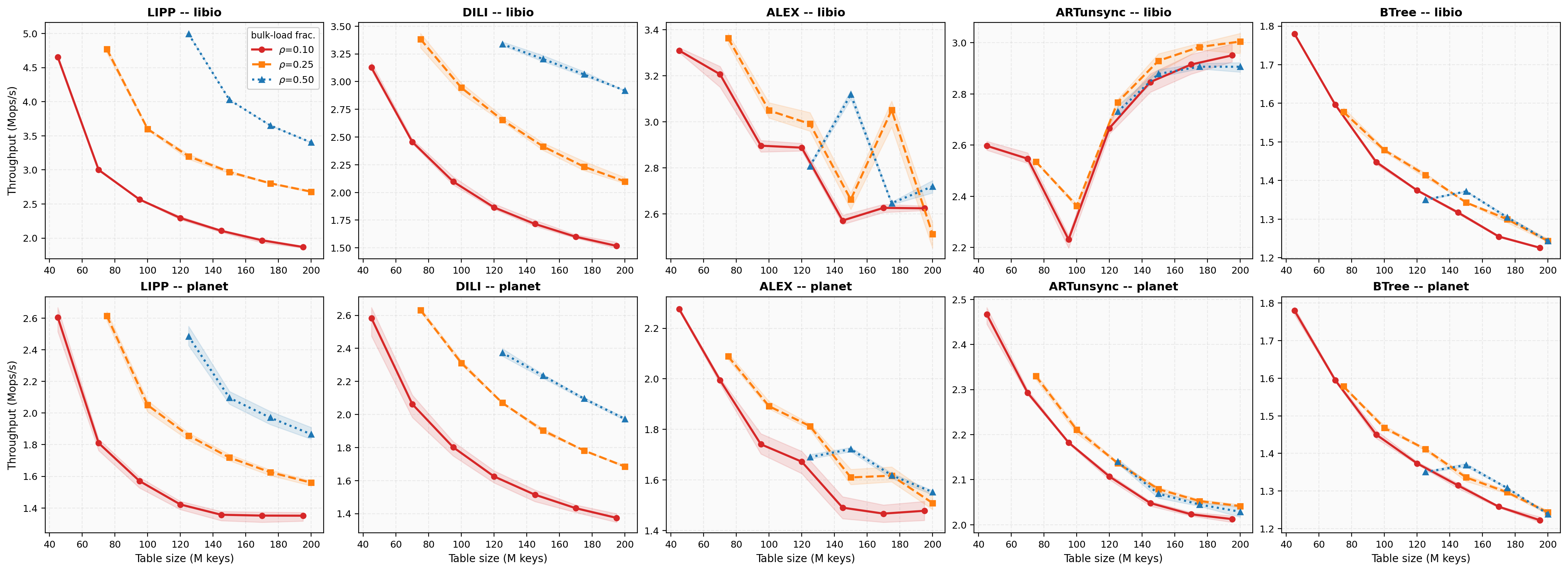}
  \Description{Ten line charts of throughput against table size, one per index per
  dataset, each showing three curves for bulk-load fractions 0.10, 0.25 and 0.50. The
  LIPP and DILI panels show three clearly separated curves, while the ALEX, ARTunsync and
  BTree panels show three curves lying almost on top of each other.}
  \vspace{-3mm}
  \caption{Throughput against table size at three bulk-load fractions, on \textbf{libio}
           (top row) and \textbf{planet} (bottom row), mean across seeds 1866, 5 and 72
           with shaded bands for the range across seeds. The runs end at 195\,M (0.10) or
           200\,M keys (0.25 and 0.50) and differ mainly in how many keys arrived by insert
           rather than by bulk load (90\%, 75\% or 50\%). Separation between the curves at
           the same table size is therefore the cost of having arrived by insert. LIPP and
           DILI (left) separate widely, while ALEX, ARTunsync and BTree (right) nearly
           coincide.}
  \label{fig:rho-paths}
\end{figure*}

The sweeps so far vary what the workload does. This one varies how the table was built.
We keep the dataset at 200\,M keys and sweep only the bulk-load fraction, so
$\rho_\text{init} \in \{0.10, 0.25, 0.50\}$ bulk-loads 20\,M, 50\,M or 100\,M keys and
the rest arrive as inserts.
The 0.25 and 0.50 runs finish holding all 200\,M keys. The 0.10 run has a 180\,M insert
pool, which fills seven whole batches of 25\,M, so it finishes at 195\,M.

Retention is the wrong statistic here, because each run covers a different size range,
so we compare throughput at the same table size instead.
Vertical separation between the three curves at the same point on the horizontal axis is
the cost of having arrived by insert.

\textbf{For LIPP and DILI that cost is large, and for the other indexes it is small.}
At the end of the run on libio, LIPP runs at 1.87\,Mops/s when 90\% of the keys were
inserted and 3.40 when only 50\% were, a spread of 82\%.
DILI spreads 92\%, from 1.52 to 2.92.
ALEX spreads 8\%, ARTunsync 3\% and BTree 1\%.
Planet gives the same ordering with smaller numbers, 38\% for LIPP and 44\% for DILI
against 5\%, 1\% and 2\% for the other three.

This is the cleanest separation in the paper between degradation caused by inserting and
degradation caused by holding more keys.
Everywhere else the two move together, because an index that has taken more inserts is
also holding more keys.
Here the final tables are nearly the same, mainly the history differs, and LIPP and
DILI end up running roughly half as fast. Periodic rebuilding points the same way
(\S\ref{sec:reindex}), by removing the history and recovering the throughput.

\textbf{It also limits the honeymoon framing.}
For LIPP and DILI a larger bulk load does not simply postpone the decline.
Across the whole range where the three runs overlap, the run that bulk-loaded more is
faster at every table size, so the benefit on the measured curve is not only a longer
peak.
ALEX does not show that separation. Its three curves nearly coincide. A larger $N_0$
moves the key count where the expansion schedule places the first burst
(\S\ref{sec:repair-loop}).

\subsection{Update- and Delete-Inclusive Workloads}
\label{app:mixedops}
\label{sec:update-inclusive}

\begin{figure*}[!t]
  \centering
  \includegraphics[width=\textwidth]{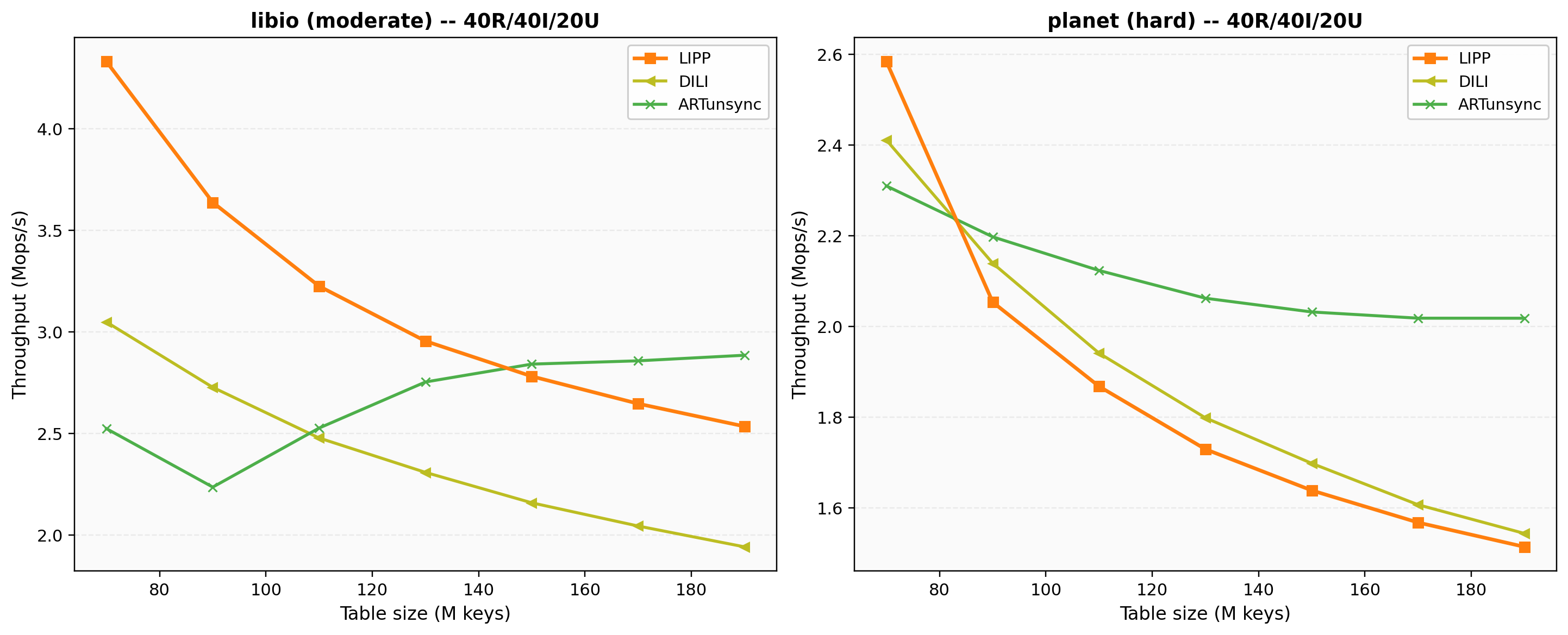}
  \Description{Two line charts showing mean throughput across three seeds vs table size,
  for LIPP, DILI, and ARTunsync under a 40 percent read / 40 percent insert / 20 percent
  update workload on libio and planet.}
  \vspace{-3mm}
  \caption{Per-batch throughput under a 40\%~read / 40\%~insert / 20\%~update workload
           on \textbf{libio} (left) and \textbf{planet} (right), mean
           across seeds 1866, 5, and 72.
           Indexes are restricted to those with verified-correct update semantics
           (see text), which are ARTunsync, LIPP and DILI.
           On libio, LIPP retains 0.585 and DILI 0.637 while ARTunsync improves to
           1.143. On planet, LIPP retains 0.586 and DILI 0.640, almost the same as on
           libio, while ARTunsync declines to 0.874.
           LIPP falls behind ARTunsync on both datasets, earlier on planet (by 90M keys)
           than on libio (about 145M).}
  \label{fig:update-inclusive}
\end{figure*}

The protocol so far is growth-only, using reads and inserts alone, and a fair question
is whether the result depends on that.
Before adding updates and deletes we audited each competitor's \texttt{update},
\texttt{remove} and \texttt{scan} path, and found several that are stubbed, semantically
wrong, or report success regardless of outcome.
One combined mix across all indexes would report incorrect numbers for those indexes,
so we run two narrower workloads instead, each restricted to the indexes whose
semantics we verified for the operation it adds.
This subsection covers updates and \S\ref{sec:delete-inclusive} covers deletes.
Only ALEX and LIPP have a correct scan, which is too narrow for a learned against
traditional comparison, so we report no scan workload.

Under a 40/40/20 read/insert/update mix, restricted to the three single-threaded indexes
with verified-correct update semantics,
\textbf{the ranking reversal persists on both datasets}
(Figure~\ref{fig:update-inclusive}). On libio, LIPP opens with a
commanding lead over ARTunsync (4.33 vs.\ 2.52\,Mops/s) and finishes behind it (2.53 vs.\
2.89) at 190M keys. On planet the same reversal happens faster. LIPP's early lead (2.58
vs.\ 2.31) is gone by 90M keys, and it finishes well behind (1.51 vs.\ 2.02). So the
reversal does not depend on a workload whose only writes are inserts.
\subsubsection*{Delete-inclusive mix}
\label{sec:delete-inclusive}

\begin{figure*}[!t]
  \centering
  \includegraphics[width=\textwidth]{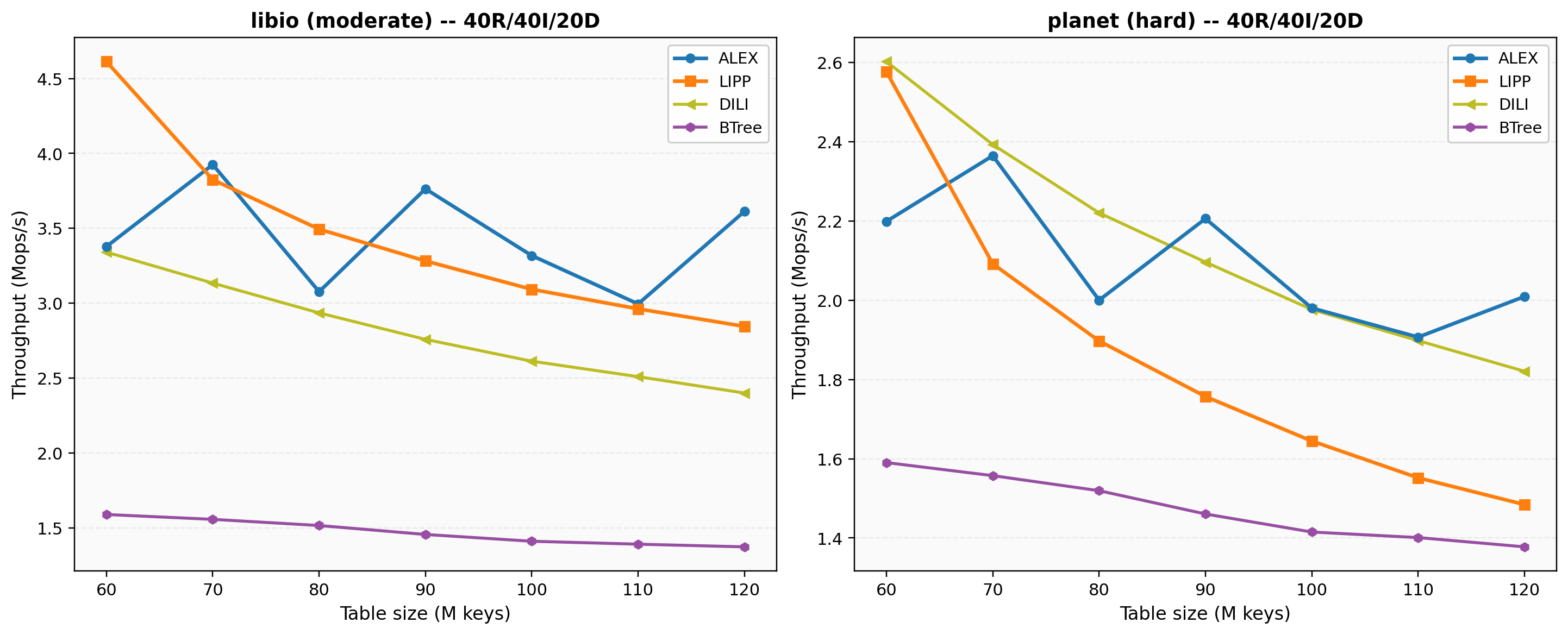}
  \Description{Two line charts showing mean throughput across three seeds vs table size,
  for ALEX, LIPP, DILI, and BTree under a 40 percent read / 40 percent insert / 20 percent
  delete workload on libio and planet.}
  \vspace{-3mm}
  \caption{Per-batch throughput under a 40\%~read / 40\%~insert / 20\%~delete workload
           on \textbf{libio} (left) and \textbf{planet} (right), mean
           across seeds 1866, 5, and 72.
           Indexes are restricted to those with verified-correct delete semantics
           (\S\ref{sec:update-inclusive}), which are ALEX, LIPP, DILI and BTree. ARTunsync
           is excluded because of the delete correctness bug described below.
           On libio, LIPP retains 0.617 and DILI 0.718 while ALEX improves to 1.070
           and BTree retains 0.864. On planet, LIPP retains 0.576 and DILI 0.700, almost
           the same as on libio, while ALEX (0.914) and BTree (0.866) again decline only
           mildly.
           ALEX shows the same drop and recover sawtooth as in the 800M scale
           extension (\S\ref{sec:scale-800m}) rather than a smooth trend.}
  \label{fig:delete-inclusive}
\end{figure*}

Figure~\ref{fig:delete-inclusive} answers the same question as
\S\ref{sec:update-inclusive} for deletes instead of updates. Under a 40/40/20
read/insert/delete mix, restricted to the four single-threaded indexes with
verified-correct delete semantics, \textbf{LIPP and DILI again lose more than the
traditional baseline, and LIPP's early lead disappears}. This mix has no learned
against traditional reversal, because its only traditional index, BTree, is the slowest
from the first batch. The reversals here are between learned indexes. On libio, LIPP
opens with the highest throughput of the four
(4.61\,Mops/s) and closes behind ALEX (2.84 vs.\ 3.61) by 120M keys. On planet, LIPP
again opens highest (2.58) and closes lowest of the learned indexes (1.48 vs.\ ALEX's
2.01); DILI's early lead over ALEX on planet (2.60 vs.\ 2.20) is also gone by the last
batch (1.82 vs.\ 2.01). BTree, the only traditional index in this subset, declines
mildly and consistently on both datasets (0.864/0.866), the same growth-only floor
behaviour as \S\ref{sec:stats}. ALEX shows the same sawtooth as in its other runs, so
its retention above 1 on libio should be read alongside the figure, not in place of
it.

ARTunsync is excluded from this figure. An exploratory delete-inclusive run surfaced a
genuine correctness fault, separate from the known success-reporting stub. Once a delete
has touched the tree, a small fraction of subsequent reads and updates return not-found
for keys that are still present (2--20 misses per 15M-op batch). Inserts remain correct
and deletes always report success by construction, so this was invisible until results
were checked against expected key state. None of ALEX, LIPP, DILI or BTree show it.

\end{document}